\documentclass[aps,pra,floats,floatfix,linenumbers,twocolumn]{revtex4-1}

\RequirePackage{lineno} 
\usepackage{soul}
\usepackage{float}
\usepackage{graphicx}
\usepackage{epstopdf}
\graphicspath{{./}{./Figs/}}
\usepackage[percent]{overpic} % overlay an image on existing figure
\usepackage{latexsym}
\usepackage{amsmath}
\usepackage{amssymb}
\usepackage{bm}
\usepackage{wasysym}

\usepackage{ifthen}
\usepackage[usenames, dvipsnames]{color}
\usepackage[colorlinks,allcolors=blue,bookmarks=true]{hyperref}

\newcommand{\eexp}[1]{\mathrm{e}^{#1}}

\newcommand{\ket}[1]{\left| #1 \right\rangle}
\newcommand{\braket}[1]{\left\langle #1 \right\rangle}
\newcommand{\Braket}[2]{\left\langle #1 \middle| #2 \right\rangle}
\newcommand{\BraKet}[3]{\left\langle #1 \middle| #2 \middle| #3 \right\rangle}

\newcommand{\avg}[1]{\left\langle #1 \right\rangle}

\newcommand{\abs}[1]{|\! #1 \!|}

\newcommand{\beq}{\begin{eqnarray}}
\newcommand{\eeq}{\end{eqnarray}}

\newcommand{\Eq}[1]{\textcolor{blue}{{Eq.}\!~(\ref{#1})}}

\newcommand{\Fig}[1]{\textcolor{blue}{Fig.}\!~\ref{#1}}
\newcommand{\Figure}[1]{\textcolor{blue}{Figure}~\ref{#1}}

\newcommand{\hide}[1]{}
\newcommand{\sect}[1]{{\bf #1.--}}

\newcommand{\aj}[0]{{\alpha j}}

\begin{document}
%%%%%%%%%%%%%%%%%%%%%%%%%%%%%%%%%%%%%%%%%%%%%%%%%%%%%%%%%%%%%%%%%%%%
%%%%%%%%%%%%%%%%%%%%%%%%%%%%%%%%%%%%%%%%%%%%%%%%%%%%%%%%%%%%%%%%%%%%

\title{Many-body dynamical localization and thermalization}

\author{Christine Khripkov$^{1,2}$, Amichay Vardi$^{1}$, Doron Cohen$^{2}$}

\affiliation{
\mbox{$^1$Department of Chemistry, Ben-Gurion University of the Negev, Beer-Sheva 84105, Israel} 
\mbox{$^2$Department of Physics, Ben-Gurion University of the Negev, Beer-Sheva 84105, Israel}}

\begin{abstract}
We show that a quantum dynamical localization effect can be observed in a generic thermalization process of two weakly coupled chaotic subsystems. Specifically, our model consists of the minimal experimentally relevant subsystems that exhibit chaos, which  are 3-site Bose-Hubbard units. Due to the high dimensionality of the composite 6-site system, the quantum localization effect is {\em weak} and can not be resolved merely by the breakdown of quantum-to-classical correspondence. Instead, we adopt an intrinsic definition of localization as the memory of initial conditions, which is not related to the underlying classical dynamics. We discuss the dynamics in the chaotic sea, and in the vicinity of the mobility edge, beyond which ergodization is suppressed.
\end{abstract}

\maketitle

%%%%%%%%%%%%%%%%%%%%%%%%%%%%%%%%%%%%%%%%%%%%%%%%%%%%%%%%%%%%%%%%
%%%%%%%%%%%%%%%%%%%%%%%%%%%%%%%%%%%%%%%%%%%%%%%%%%%%%%%%%%%%%%%%
\section{Introduction}

The study of {\em dynamical} localization in a low-dimensional chaotic system was pioneered in the publication ``Chaos, Quantum Recurrences, and Anderson Localization" by Fishman, Grempel and Prange \cite{Fishman}, which had been motivated by the puzzling numerical observation of quantum-suppressed chaotic diffusion, by Casati, Chirikov, Izrailev, and Ford \cite{Casati}. For an extended period, most research efforts focused on the quantum kicked-rotor system, regarded as the quantized version of the Chirikov standard map (see \cite{standard_map} and references within). This system has one degree of freedom, and the driving (kicking) effectively adds an extra half freedom. 

\sect{Dynamical localization}
In an Anderson tight-binding {\em disordered} chain all the eigenstates are exponentially localized, and therefore the spreading of a wave packet in space ($x$) is always suppressed after some break time ($t^*$). The break time is related to the localization length of the eigenstates. Similarly, in the kicked-rotor problem, the variable ``$x$" is the angular momentum. While there is no disorder in this one+half degree of freedom dynamical system, the chaotic classical dynamics of ``$x$" is stochastic-like,  provided the kicks are strong enough. Thus, in analogy to Anderson localization, the diffusive spreading is suppressed in quantum simulations after some break time $t^*$, indicating that the Floquet eigenstates are localized. The relation between the break time and the localization length can be formalized \cite{Chirikov,Shep,Dittrich,Cohen}, and both are related to the disorder strength (in the Anderson model) or to the kicking strength (in the kicked-rotor system).               

\sect{Localization in higher dimensions} 
Strong Anderson localization is not necessarily present in higher dimensions. This is explained by scaling theory and renormalization group methods. In general, there is a mobility edge that separates regions where strong localization prevails from those where spreading is unaffected by quantization. Nevertheless, in the latter case it may be possible to obtain a {\em weak}-localization effect, meaning that the probability of return to the initial region is affected. Weak localization is thus a memory effect, expressed as a dependence of the outcome on initial conditions.          

\sect{Localization in finite systems} 
Most condensed matter physics literature concerns disordered infinite-volume systems, and the weak-localization effect is associated with time-reversal symmetry. The paradigmatic kicked rotor is also formally identical to an infinite-volume chain, 
because the range of angular momentum is unbounded, and the energy is not a constant of motion.   
Still, the notion of quantum localization has been extended also to the realm of undriven finite systems that have a finite-volume energy surface. In this context we can  define localization as the absence of ergodicity. If the system is ergodic, an initial cloud of evolving points (``classical wave packet") is expected to eventually be smeared uniformly over the entire energy surface, approaching a microcanonical-like distribution. Quantum localization can therefore be defined as the lack of ergodicity in quantum systems for which the corresponding classical dynamics is ergodic. This means that even after a very long time the quantum probability distribution is not microcanonically uniform. Heller \cite{Heller} has realized that (weak) dynamical localization due to interference is generically implied by short-time classical recurrences.

\sect{Classical localization}
Idealized hard chaos is an exception; a generic finite undriven system features a structured mixed phase space that supports both chaotic and quasiregular motion. Ergodization over the entire energy surface is not guaranteed, or it might be extremely slow, involving exponentially long timescales. Consequently, the practical definition of {\em localization} becomes more subtle. If we define localization as a lack of ergodicity, then it should be clear that it may arise also in a purely {\em classical} context, due to fragmentation of the energy surface. This fragmentation can happen either due to energetic barriers between different regions (e.g., self-trapping \cite{selftrap}), or due to dynamical obstacles of the Kolmogorov-Arnold-Moser (KAM) type \cite{KAM}. 

\sect{Quantum localization}
Coming back to the definition of quantum localization, Heller \cite{Heller} has suggested defining a measure for it by comparing the quantum and the classical exploration volumes. Quantum ergodicity, in this perspective, means that the ratio of volumes is identical to the value predicted by random matrix theory (RMT): somewhat less than unity due to quantum fluctuations. Quantum localization then means that this ratio is significantly smaller than the RMT expectation.

\sect{Intrinsic localization measure}
In the realm of high-dimensional finite systems, the standard localization measures, which are based on contrasting quantum versus classical behavior, typically provide inconclusive results. Due to the mixed, complicated structure of the phase space, quantum-to-classical correspondence is \emph{a priori} poor, and is not in accordance with the predictions of RMT. Hence, simply comparing quantum to classical spreading volumes would lead to the conclusion that quantum localization always shows up. Correspondence-based measures thus trivialize the concept of quantum dynamical localization and furthermore, can not be extended to systems such as spin chains, where the classical limit is meaningless. It is therefore highly desirable to find an {\em intrinsic} measure for localization that is not prejudiced by the existence of a classical limit. Below, we adopt such a correspondence-free definition of dynamical localization.

\sect{Additional subtleties}
Additional subtleties are involved in the calculation of the standard measure of localization. 
It is impossible in practice to verify that a classical system is {\em not} ergodic. Very long simulations can be carried out, but their results cannot be trusted: on the one hand, the exploration process might involve exponentially long timescales, e.g., due to Arnold diffusion \cite{basko}; on the other hand, long simulations might be infected by numerical inaccuracies.
These complications do not arise in the quantum context, as the $t=\infty$ saturation profile of the evolving wave packet can be anticipated via diagonalization of the Hamiltonian (see Sec.~III below). Consequently, a time-dependent quantum simulation is not even required in order to determine whether the system is quantum-ergodic. The entire analysis of quantum localization can be formulated based on the dependence of known asymptotic quantum distributions on the initial preparation, without any need for dynamical simulations.

\sect{Thermalization}
The typical thermalization setup consists of two {\em weakly} coupled subsystems. In introductory statistical mechanics textbooks, one of the subsystems is assumed to be a large chaotic environment. It is then argued that the ergodization of the combined system implies equilibration of the smaller subsystem (which does not have to be chaotic). In point of fact, since chaotic ergodization necessitates at least two degrees of freedom, either one or both subsystems can be chaotic, but neither have to be large. It is then possible to define a reaction coordinate $x$ that quantifies the exchange of energy or particles between two such moderately sized subsystems, and to show that the associated dynamics obeys a Fokker-Planck equation \cite{trmPRL}. The question arises whether upon quantization the implied thermalization process is arrested due to dynamical localization.

\sect{Minimal setup}
In a recent paper \cite{Chris} we looked for a dynamical localization effect in the many-body thermalization of a minimal Bose-Hubbard model. Namely, we have considered an isolated 4-site system with interacting bosons, where three sites are 
strongly coupled (the trimer) and an additional site (monomer) is weakly coupled to them.
The combined model has 3 degrees of freedom (corresponding to  occupation differences and relative phases between the sites). The $x$ coordinate was the occupation of the trimer. 
If the dynamics is treated classically, the $x$ variable thermalizes within a region that is determined by the boundaries of the chaotic sea; i.e., the $x$ distribution obeys, as expected, a Fokker-Plank equation. By contrast, the quantum dynamics exhibits {\em dynamical} localization if $x$ starts at the peripheral regions (i.e., if the simulation is initiated with a large occupation imbalance).  
We have developed a theory for finding the mobility edge that separates the $x$ region of quantum ergodicity from the regions where localization manifests itself. For that purpose we have combined the break-time phenomenology of Chirikov and followers \cite{Chirikov,Shep,Dittrich,Cohen} with the phase space exploration phenomenology of Heller \cite{Heller}.

\sect{Generic setup}
In this work we explore dynamical localization in the more generic case of trimer-trimer thermalization. The main conceptual difference from the trimer-monomer model is that in the present case both subsystems are chaotic. The motivation for this study is related to the possible implication of such an observation: namely, manifestation of dynamical localization in the trimer-trimer system implies that it is relevant for the analysis of thermalization in larger disordered arrays, since such arrays can be viewed as chains of weakly interacting subsystems, some (or all) of which are chaotic. It should be re-emphasized that, unlike the disorder-induced many-body localization \cite{MBL,MBLb}, the cause of dynamical localization in this work is interaction-induced chaos on the microscopic scale, and the thermodynamic limit is not an issue.

\sect{Effective $\hbar$}
The Bose-Hubbard model is formally equivalent to that of coupled oscillators, and therefore one may use the traditional distinction between classical and quantum descriptions of the same system. In this context the number of particles per subsystem plays the role of inverse $\hbar$; namely, proper scaling of the dynamical variables gives an effective dimensionless Planck constant ${\hbar_{\text{eff}}=1/N_{\text{subsystem}}}$.

\sect{The localization question}
By now the manifestation of dynamical localization in low-dimensional chaos, such as the kicked-rotor, is hardly surprising. But as we go up in dimensionality the observation of a quantum localization effect becomes questionable. The break-time phenomenology suggests quantum localization whenever the classical exploration becomes slow in some sense (see \cite{Chris}), but the argument is rather speculative and has never been tested for high-dimensional chaos, except for the trimer-monomer system. 
It has to be realized that the trimer-trimer system is much more challenging for dynamical localization because the peripheral regions of the chaotic sea are vast, unlike those of the trimer-monomer system.   
Indeed, as shown below, dynamical localization in a trimer-trimer system is a rather weak effect that requires a refined numerical procedure for its detection. A major challenge is to distinguish a ``novel type" of dynamical localization that should not be confused with energetic metastability (self-trapping), with perturbative localization, or with classical localization due to KAM structures. This novel type of dynamical localization appears within the chaotic sea and cannot be explained in a simple way.

%%%%%%%%%%%%%%%%%%%%%%%%%
\sect{Outline}
In Sec.~\ref{model} we introduce the model Hamiltonian, characterize the trimer as a subsystem, and define the required representation for the discussion of the thermalization process. Numerical demonstration of dynamical localization is provided in Sec.~\ref{numerical}. The ergodicity measure whose objective is to identify the mobility edges of the chaotic sea is defined in Sec.~\ref{measures}; it is used for the analysis of classical localization in Sec.~\ref{classical} and of quantum localization in Sec.~\ref{quantum}. Finally, in Sec.~\ref{conclusion} we emphasize the nontriviality of the observed dynamical localization, as opposed to perturbative localization on the one hand, and semiclassical nonergodicity on the other.

%%%%%%%%%%%%%%%%%%%%%%%%%%%%%%%%%%%%%%%%%%%%%%%%%%%%%%%%%%%%%%%%
\section{The model}
\label{model}

Our building blocks are a pair of Bose-Hubbard trimers, labeled by $\alpha=L,R$.  Each trimer is described by the Hamiltonian  
\beq\label{eq:Hamiltonian_alpha}
\mathcal{H}_{\alpha} \ = \ \frac{U}{2}\sum_{j=1}^{3} \hat{n}_\aj^2 -\frac{K}{2}\sum_{j=1,3}\left(\hat{a}_\aj^\dagger \hat{a}_{\alpha 2} + \text{H.c.} \right)~.
\eeq
The operators $\hat{a}_\aj^\dagger$ and $\hat{a}_\aj$, respectively, create or annihilate a particle on site $\alpha j$, while $\hat{n}_\aj=\hat{a}_\aj^\dagger \hat{a}_\aj$ counts particles in this site. The parameter $K$ is the tunneling strength between neighboring sites, while $U$ describes interaction between two particles. In the absence of intertrimer coupling (as well as for our particular choice of coupling; see below), the populations $N_\alpha=\sum_j \hat{n}_{\alpha j}$ are separately conserved. The total number of particles is $N=N_L+N_R$. In our simulations we assume for simplicity that ${N_L=N_R=N/2}$.

In order to simulate a thermalization process, the two trimer subsystems are weakly coupled. For this purpose we employ a nearest-neighbor interaction of particles that occupy the same location,
\begin{align}
\label{eq:Hamiltonian_parts}
\mathcal{H} &= \mathcal{H}_L+\mathcal{H}_R + \mathcal{H}_c~,\\
\label{eq:Hamiltonian_coupling}
\mathcal{H}_c &= \frac{V}{2}\sum_{j=1}^{3} \hat{n}_{Lj}\,\hat{n}_{Rj}~,
\end{align}
with $V$ denoting the nearest-neighbor interaction strength. This extended Bose-Hubbard model \cite{Dutta15,Dey17} can be realized using the long-range dipole-dipole interaction between lattice-BECs of particles with electric or magnetic dipole moments \cite{Lahaye09,Trefzger11}.

The dimensionless parameters of the model are
\beq
u &\equiv& \frac{NU}{K} \ \ = \ \ \text{(intra-trimer nonlinearity)} \\
v &\equiv& \frac{NV}{K} \ \ = \ \ \text{(inter-trimer coupling)} 
\eeq
Without loss of generality we set the units of time such that the hopping frequency is ${K=1}$. This holds for all the numerical results that will be presented. 

In what follows, we refer to the exact many-body evolution under the Hamiltonian of \Eq{eq:Hamiltonian_parts} as the {\em quantum dynamics}. Its large-$N$ classical limit is obtained \cite{meanfield} by replacing the field operators $\hat{a}_\aj$ by $c$-numbers ${a_\aj=\sqrt{N_\alpha I_\aj}\eexp{i\varphi_\aj}}$, where the normalized occupations ${I_\aj\in[0,1]}$ and the phases ${\varphi_\aj\in[0,2\pi)}$ are canonical action-angle variables. 
Since the bosonic Hamiltonian is formally equivalent to that of a coupled-oscillators system, one may use the traditional distinction between classical and quantum descriptions of the same system \cite{meanfield}. 
The commutation relations imply that the dimensionless Planck constant is ${\hbar_{\text{eff}} = 1/N_{\alpha}=2/N}$.
The classical form of the Hamiltonian is obtained formally in a straightforward manner from the second-quantized Hamiltonian using the above-defined substitutions, and the prescription ${ \mathcal{H}^{cl} = \hbar_{\text{eff}} \mathcal{H} }$. This leads to 
\beq
\label{eq:Energy_total}
\mathcal{H}^{cl} = \mathcal{H}_{L}^{cl} + \mathcal{H}_{R}^{cl} + \frac{v}{4}\sum_{j=1}^{3} I_{Lj}\,I_{Rj}~,
\eeq
with 
\beq
\label{eq:Energy_alpha}
\mathcal{H}_{\alpha}^{cl} = \frac{u}{4}\sum_{j=1}^{3} I_\aj^2 -\sum_{j=1,3}\sqrt{I_\aj I_{\alpha 2}}\cos(\varphi_\aj-\varphi_{\alpha 2})~.
\eeq
Thus we have only two dimensionless parameters $u$ and $v$ that determine the (scaled) classical dynamics, 
while in the quantum version we have the additional dimensionless parameter $\hbar_{\text{eff}}$.

%%%%%%%%%%%%%%%%%%%%%%%%%%%%%%%%%%%%%%%%
%%% Figure %%%%%%%%%%%%%%%%%%%%%%%%%%%%%
\begin{figure}
\centering 
\begin{overpic}[width=1\linewidth]{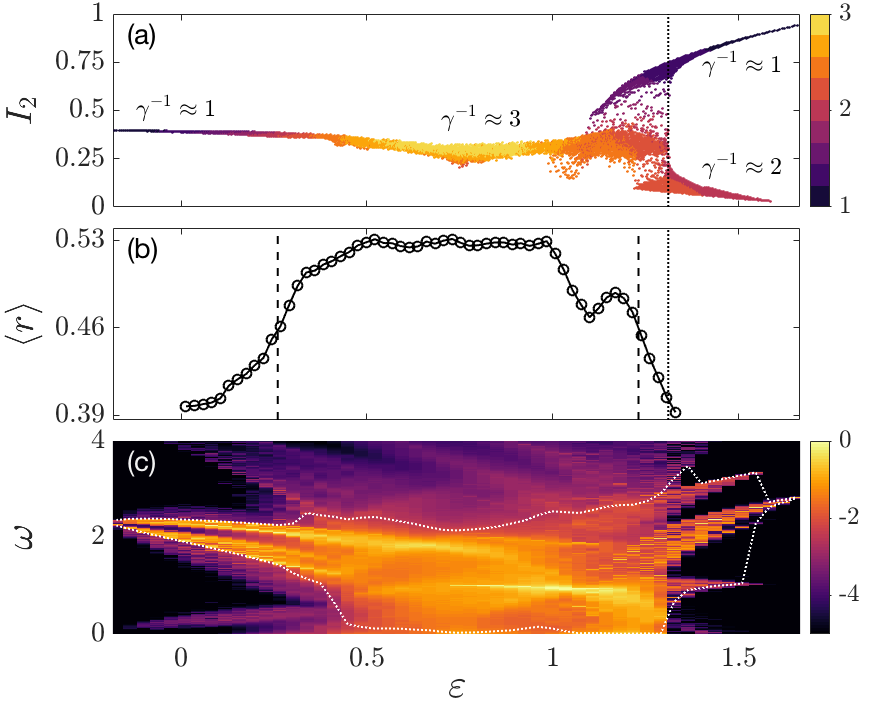}
\put(38,71){\includegraphics[width=0.22\linewidth]{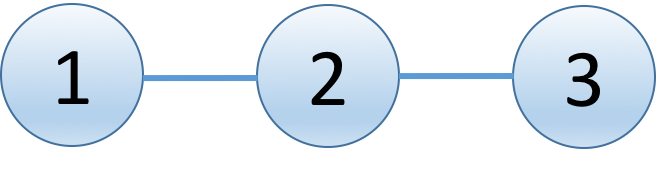}}  
\end{overpic}
\caption{\label{f:trimer} 
(Color online) The spectrum of an isolated trimer. 
\textbf{Panel (a)}: Each point represents an eigenstate $\ket{\varepsilon}$ for ${N_\alpha=350}$ particles and ${u=6}$, positioned according to its energy~$\varepsilon$, and the normalized expectation value of the middle-site occupation~$I_2$. The points are color-coded by inverse purity~$\gamma^{-1}$. 
Inset: The isolated trimer.  
\textbf{Panel (b)}: The locally averaged level spacing correlation function $\avg{r}$ for ${N_\alpha=450}$. The chaotic range is defined as the energies for which $\avg{r}$ lies above the value ${(\avg{r}_\text{Poisson}{+}\avg{r}_\text{GOE})/2=0.458}$ that distinguishes between GOE and Poisson statistics.  
Vertical lines indicate the lower and upper chaos borders (dashed) and the self-trapping threshold (dotted).
Panel \textbf{(c)}: Power spectra of the trimer occupation $I_2(t)$, color-coded using a $\log_{10}$ scale. Each column is the average spectrum for a different value of $\varepsilon$. Dotted white boundaries enclose $97$\% of the total power.  
}\end{figure}
%%% End Figure %%%%%%%%%%%%%%%%%%%%%%%%%

%%% old placement for Fig2!

%%%%%%%%%%%%%%%%%%%%%%%%%%%%%%%%
\subsection{The trimer}

The chaotic dynamics of at least one subsystem is a necessary ingredient in thermalization. The characterization of chaos for a single trimer is therefore essential before considering the coupling between two such subsystems. The detailed description of the trimer and the methodology of chaos determination are deferred to Appendices A and B, respectively. Our main objective is to set the range of energies in which the trimer is classically chaotic. The straightforward way to identify chaos is to run classical simulations throughout phase space for each value of the energy. This procedure is inefficient and time-consuming. However, it is well established that classical chaos can be determined from the level spacing statistics of the quantum spectrum \cite{HaakeTextbook}. Thus, via one diagonalization of the quantum Hamiltonian we obtain the complete  information on the chaotic energy range for any value of the characteristic parameters. 
From the same diagonalization we can extract additional information on the underlying chaos by considering the quantum eigenstates. According to the eigenstates thermalization hypothesis (ETH), a somewhat misleading term in the present context, the eigenstates that dwell within the classically chaotic phase space regions are expected to be ergodically spread throughout them \cite{Olshanii}. This ergodicity is reflected in the expectation value of observables, which attain their thermal values.        

Moderate-strength interactions in the trimer give rise to chaos. However, stronger interaction may also cause {\em self-trapping}, which is localization within an isolated part of the energy surface that is surrounded by a (higher or lower energy) forbidden region. The splitting of the energy surface into separate, disconnected regions can be reflected in a splitting in the distribution of the site occupation. 

When all the particles are condensed within a single one-particle state (either a site state or some orbital that constitutes a superposition of site states), the resulting many-body state is known as a ``coherent state";  it is represented by a Gaussian-like distribution in phase space.The spread of the eigenstate in phase space can be measured by its purity (see definition in Appendix~A). The inverse purity $\gamma^{-1}$ estimates the number of one-particle states that participate in the formation of the many-body state. High purity of order unity implies that only a single state is occupied. Low purity implies that the occupation is fragmented; for a trimer eigenstate the lowest purity is~$1/3$ (as there are three independent one-particle states). 

The spectrum of an isolated trimer (say, ${\alpha=R}$) is illustrated in \Fig{f:trimer}(a) for ${u=6}$. 
Each point represents an eigenstate $\ket{\varepsilon_R}$ of the Hamiltonian $\mathcal{H}_{R}$, positioned horizontally according to its eigenenergy ${\varepsilon \equiv \braket{\mathcal{H}_{R}}/N_R}$, and vertically according to the normalized expectation value of the middle-site occupation, ${I_2 \equiv \braket{\hat{n}_2} / N_R}$. 
The split of the distribution of occupations in \Fig{f:trimer}(a) provides a clear numerical identification of the self-trapping border, indicated by a vertical dotted line.

The points in \Fig{f:trimer}(a) are color-coded by the inverse one-particle purity of the eigenstates $\gamma^{-1}$. Detailed analysis of its implications is provided in Appendix~A. In the chaotic energy region one expects low purity, because eigenstates are spread over large chaotic regions of the energy surface. But the precise determination of the chaotic region requires further effort. \Fig{f:trimer}(b) displays the $r$ measure of level-spacing statistics (see Appendix~B for definition and a detailed analysis). The value ${r \approx 0.530}$ indicates RMT statistics, the Gaussian orthogonal ensemble (GOE) in particular, that is expected for a chaotic repulsion of levels in time-reversal-invariant systems. This is opposed to the value ${r \approx 0.386}$ that indicates Poissonian statistics, innate to quasi-integrable regions. Consequently, we identify the chaotic range ${0.26 < \varepsilon < 1.23}$, based on the halfway mark ${\avg{r}>(0.386+0.530)/2}$. 

The chaotic range has been further verified by direct classical phase space analysis as in \cite{trmPRL}, inspecting the Poincar\'e sections generated by trajectories at different energies. Here we provide another piece of proof that will become useful in later analysis. 
In the classical limit, the Hamiltonian \Eq{eq:Energy_alpha} describes the dynamics of coupled nonlinear oscillators. From it we can derive the classical equations of motion for the occupations, and then propagate them. For motion on an energy surface ${\mathcal{H}^{cl}_\alpha(I_{1,2,3},\varphi_{1,2,3})=\varepsilon}$ we can obtain classical power spectra of the site occupations (see Appendix~C). This is shown in \Fig{f:trimer}(c) for the central trimer site $I_2$; the picture for $I_{1,3}$ is qualitatively similar. Within the central energy region the classical motion has a wide frequency content, which is the hallmark of chaos (see Appendix~B).

%%% Figure %%%%%%%%%%%%%%%%%%%%%%%%%%%%%
\begin{figure}\centering 
\begin{overpic}[width=1\linewidth]{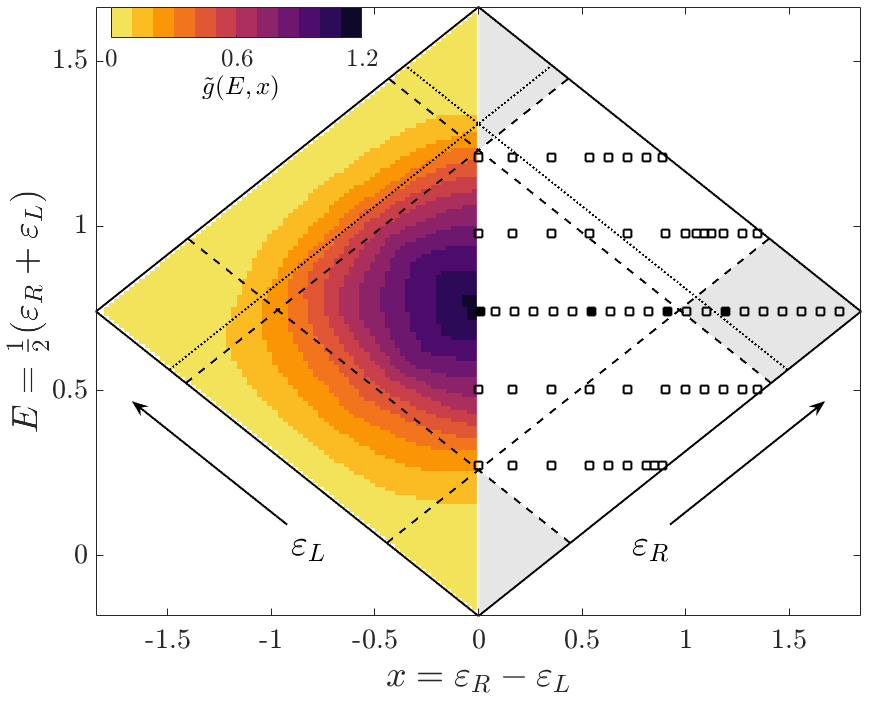}
\put(75,65){\includegraphics[width=0.18\linewidth]{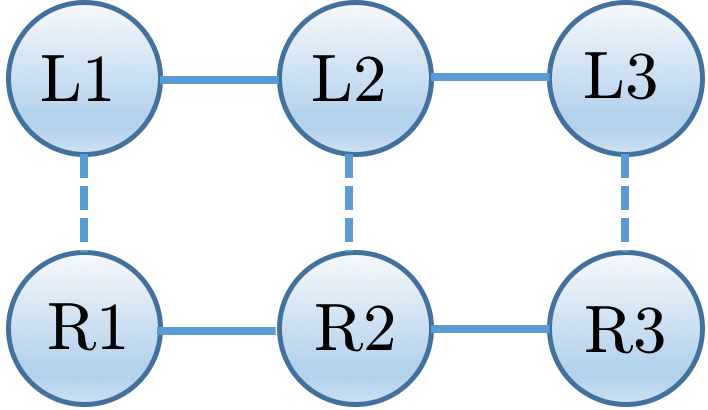}}  
\end{overpic}
\caption{\label{f:doubleTrimer}  
Illustration of the spectrum for the noninteracting (${v=0}$) double trimer with ${u=6}$. 
Each direct product eigenstate of ${\mathcal{H}_L+\mathcal{H}_R}$ is represented by a point in the $(E,x)$ plane. For a large $N$ these points form a dense lattice within the classically allowed range (marked by the black solid boundary).  
The diagonal dashed lines indicate the lower and upper chaos borders; the dotted lines indicate the self-trapping threshold. Within the gray regions both trimers are quasiregular.
Markers show the initial locations of the semiclassical simulations of \Fig{f:localizationSummary}; filled markers further refer to the selected simulations in \Fig{f:distributions}. 
The color in the left half of the figure indicates the value of the joint density of states.
{\em Inset:} The configuration of the double-trimer system. Adjacent sites within a trimer are strongly coupled (solid lines), while sites that belong to different trimers are weakly coupled (dashed lines).} 
\end{figure}
%%% End Figure %%%%%%%%%%%%%%%%%%%%%%%%%

%%%%%%%%%%%%%%%%%%%%%%%%%%%%%%%%
\subsection{Two uncoupled trimers}

Considering the two trimers together, yet without interaction, we obtain the energy landscape illustrated in \Fig{f:doubleTrimer}. Each point represents an eigenstate $\ket{E,x}$ of ${\mathcal{H}_{R}+\mathcal{H}_{L}}$, positioned according to 
\beq
E &=& \frac{1}{2}(\varepsilon_R+\varepsilon_L)~, \\
x &=& (\varepsilon_R-\varepsilon_L) ~.
\eeq
Note again that scaled units are used for these variables; i.e., the hopping frequency is set to ${K=1}$, and the trimer eigenenergies are normalized per particle.
Note also that in our previous work on the thermalization of a trimer-monomer system \cite{Chris}, particle exchange was allowed and $x$ was the occupation difference. The physics of dynamical localization does not depend on the nature of $x$.

Let us label the single trimer density of states by $g(\varepsilon)$. Considering two uncoupled trimers, their joint density of states $\tilde g(E_0;x)$ with respect to $x$, within an energy shell $(E_0,E_0+dE)$, is given by the product ${g(E_0+x/2) \, g(E_0-x/2)dE}$, indicated by color in \Fig{f:doubleTrimer}. An ergodic distribution is defined as the microcanonical uniform occupation of all states. Therefore, up to a normalization constant, this distribution is
\beq
P_{\text{erg}}(x) \ \ \propto \ \ g(E_0+x/2) \, g(E_0-x/2) ~.
\label{perg}
\eeq

%%%%%%%%%%%%%%%%%%%%%%%%%%%%%%%%%%%%%%%%%%%%%%%%%%%%%%%%%%%%%%%%%%%%%%%%%%%%%%%%%%%%%%%%%%%%%%
\section{Spreading and Ergodization}
\label{numerical}

Below we describe the dynamics in terms of a probability distribution~$P_t(x)$. For an idealized thermalization process of coupled chaotic subsystems one expects this function to obey a Fokker-Plank equation \cite{trmPRL}, leading to thermalization such that $P_{\infty}(x) \sim P_{\text{erg}}(x)$. Here we are dealing with a complicated nonhomogeneous mixed chaotic dynamics; therefore we cannot expect a simple diffusive behavior. Still we do expect that the spreading will look quasistochastic, meaning that the spreading has the same rate if we plot it against the scaled time $\tilde{t} = v^2t$. This should be contrasted with ballistic transport, for which ${\text{var}(x) \sim v t}$. The quasistochastic time dependence is valid classically whenever the ``diffusion" along~$x$ is slow compared to the ergodization within the  subsystems (this is the weak-coupling assumption). In the quantum case it can be regarded as a Fermi-golden-rule picture (FGR). We shall explain below that for very small $v$ the FGR breaks down, leading to trivial perturbative localization, which is not of much interest for us. Our focus is on couplings $v$ that can be regraded as classically weak, but quantum mechanically large. This is the regime where semiclassical perspective is most appropriate, and this is also the regime where dynamical  localization effects are not trivial.

%%%%%%%%%%%%%%%%%%%%%%%%%%%%%%%%%%%%%%%%%%%%%%%%%%
\subsection{Quantum dynamics of coupled trimers}

Preparing the double-trimer system in one of the coupling-free eigenstates  $\ket{E_0, x_0}$ described above, its exact quantum dynamics is analyzed via the time-dependent probability distribution  
\beq\label{eq:P_x}
P_t(E,x|E_0,x_0) \ = \ \abs{\BraKet{E,x}{\eexp{-i\mathcal{H} t}}{E_0, x_0}}^2~.
\eeq
The $x$ range is divided into bins of arbitrary, but small width $\delta x$. For a given initial state $\ket{E_0, x_0}$
we define the coarse-grained distribution
\beq
P_t(x) \ = \ \sum_{x<x'<x+\delta x} P_t(E,x'|E_0,x_0)~.
\eeq
Finally, we can average over a sufficiently long time  
to obtain a relatively smooth saturation profile.
Formally we write
\beq\label{eq:average_profile}
P_{\infty}(x) \ = \ \lim_{t\rightarrow\infty} \frac{1}{t} \int_0^t P_{t}(x) dt~.
\eeq

In fact, the exact saturation profile can be obtained directly from the eigenstates of the Hamiltonian. 
Using the short notation $\ket{X} \equiv \ket{E,x}$, and expanding each state in the basis $\ket{\mathcal{E}_n}$ of the full Hamiltonian $\mathcal{H}$, the time-dependent distribution takes the form
\beq \nonumber
P_t(X) = 
\sum_{n,m} \Braket{X}{\mathcal{E}_n}\!\Braket{\mathcal{E}_n}{X_0}\! \Braket{X_0}{\mathcal{E}_m}\! \Braket{\mathcal{E}_m}{X}  
\eexp{i(\mathcal{E}_m-\mathcal{E}_n)t}~.
\eeq
For a nondegenerate spectrum, upon time averaging, the oscillating terms cancel out and one obtains the result ${P_{\infty}(X)= P_{\text{sat}}(X)}$, where   
\beq\label{eq:sat_profile}
P_{\text{sat}}(X)  = \sum_n |\Braket{X}{\mathcal{E}_n}|^2 \, |\Braket{\mathcal{E}_n}{X_0}|^2~.
\eeq
In the following, the term \emph{quantum saturation profile} shall always refer to the exact result of \Eq{eq:sat_profile}.

%%%%%%%%%%%%%%%%%%%%%%%%%%%%%%%%%%%%%%%%%%%%%%%%%%%%%%%%%%%%%%%%%%%%
\subsection{Semiclassical dynamics of coupled trimers} 

In the semiclassical picture, a quantum wave packet is represented by a ``cloud" of phase space points initially residing on the energy surface $(E_0,x)$ (see Appendix~C). This approach is similar to the truncated Wigner approximation. Coupling the two trimers causes the cloud to spread within the accessible energy shell. For weak intertrimer coupling, if the dynamics is strictly ergodic within the energy shell, the distribution $P_{\text{erg}}(x)$ of \Eq{perg} is obtained after a sufficiently long time, regardless of the details of the initial preparation. 

The semiclassical picture requires a large number of time-dependent simulations. We assume an initial preparation that is represented by a microcanonical cloud of 1000 points, spread uniformly over $I_{\alpha j}$ and $\varphi_{\alpha j}$ values corresponding to ${(E_0,x_0)}$ as determined by the {\em unperturbed} Hamiltonians $\mathcal{H}^{cl}_\alpha$ of uncoupled trimers. The full numerical procedure is described in Appendix~C. Individual points of the cloud are propagated under the canonical equations of motion ${d b_\aj/dt=-i\partial \mathcal{H}^{cl}/\partial b_\aj^*}$, where ${b_\aj\equiv a_\aj/\sqrt{N_\alpha}=\sqrt{I_\aj}\eexp{i\varphi_\aj}}$.  The final distribution of points $P_t(x)$ is obtained by counting the number of points falling within each bin $(x,x+\delta x)$ at a time $t$.

%%%%%%%%%%%%%%%%%%%%%%%%%%%%%%%%%%%%%%%%%%%%%%%%%%%%%%%%%%%%%%%
\subsection{Quantum versus semiclassical simulations}

\begin{figure}
\centering 
\includegraphics[width=1\linewidth]{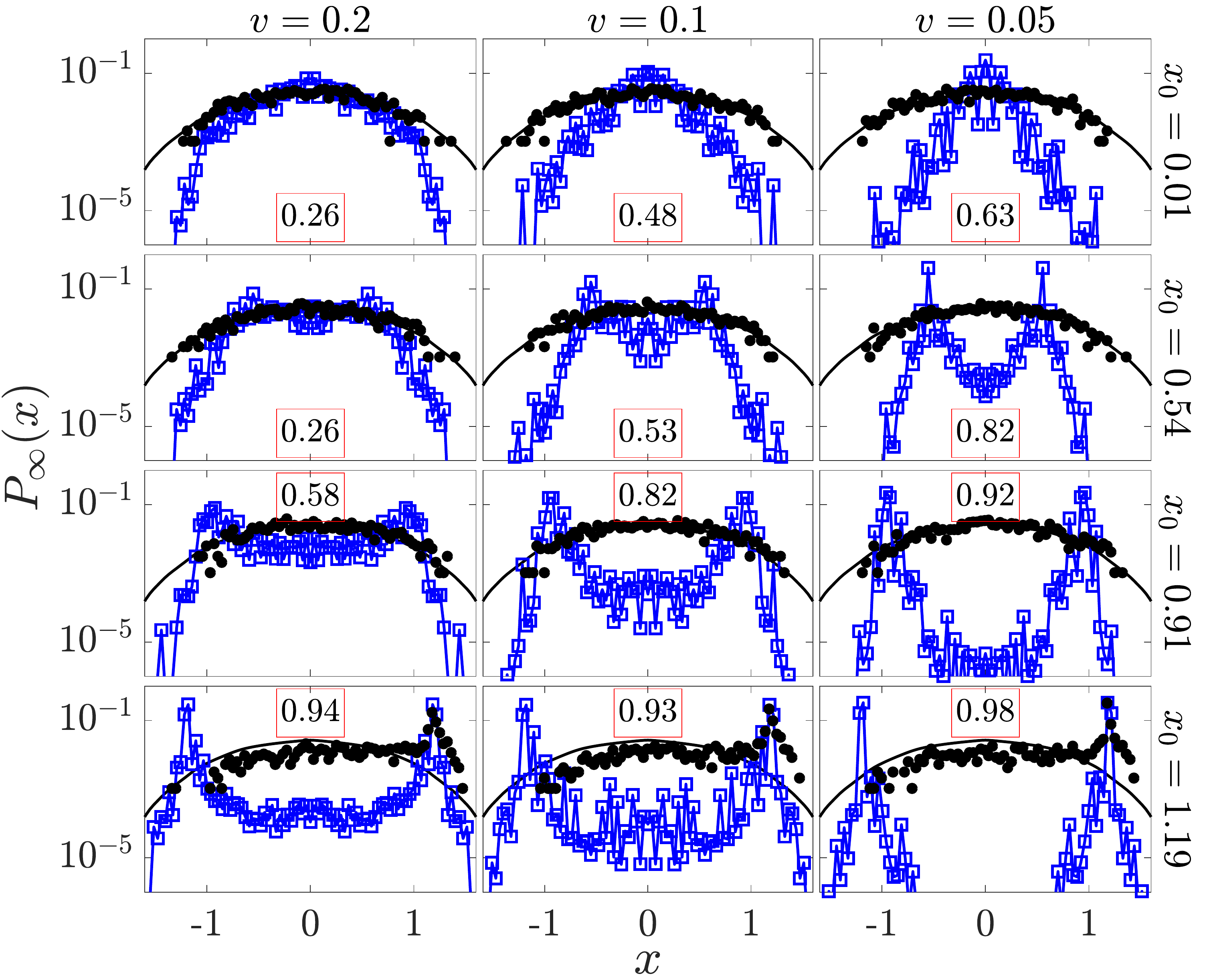}  
\caption{\label{f:distributions}  
(Color online) Semiclassical long-time probability distributions $P_\infty(x)$ evaluated at $\tilde{t}=450$ (black \text{\large $\circ$}), and quantum saturation profiles $P_\text{sat}(x)$ (blue $\square$), compared with the ergodic distribution $P_\text{erg}(x)$ (solid black line).
The initial energy is $E_0=0.74$; the value of $x_0$ varies with each row, while the value of $v$ varies with each column. The initial conditions are marked by filled symbols in \Fig{f:doubleTrimer}. The boxed number in each panel indicates the value of $f_\infty$ evaluated from \Eq{eq:ergodicity} for the  quantum saturation profiles.
In all figures semiclassical simulations are for a cloud of 1000 randomly selected phase space points, initially uniformly covering the energy surface $(E_0,x_0)$. Quantum simulations correspond to direct product states with $N_L=N_R=24$, locally averaged over several states lying in the range $E=E_0\pm 0.02$ and $x=x_0\pm 0.025$. 
}\end{figure}

\Figure{f:distributions} displays the results of representative quantum and semiclassical simulations. The initial energy, $E_0$, is the same for all simulations, while the initial trimer energy difference, $x_0$, is different for each row of panels (these initial conditions are marked in \Fig{f:doubleTrimer}). The value of  $v$ is different for each column.
The probability distribution $P_{\infty}(x)$ in the long-time limit [based on time evolution for semiclassical simulations, or on the  saturation profile of \Eq{eq:sat_profile} for quantum simulations] can be compared with the ergodic distribution $P_{\text{erg}}(x)$ that is indicated by the black line.   
We see that semiclassical simulations come close to the ergodic distribution for ${x_0=0.01, 0.54, 0.91}$ but not for  ${x_0=1.19}$, irrespective of~$v$. By contrast, quantum simulations are sensitive to~$v$, and either approach the ergodic distribution for a  more limited range of $x$ than their semiclassical counterparts, or are completely nonergodic. In the next section our objective is to quantify this observation.

%%% Figure %%%%%%%%%%%%%%%%%%%%%%%%%%%%%
%%% End Figure %%%%%%%%%%%%%%%%%%%%%%%%%

%%% Figure %%%%%%%%%%%%%%%%%%%%%%%%%%%%%
\begin{figure}
\centering 
\includegraphics[width=1\linewidth]{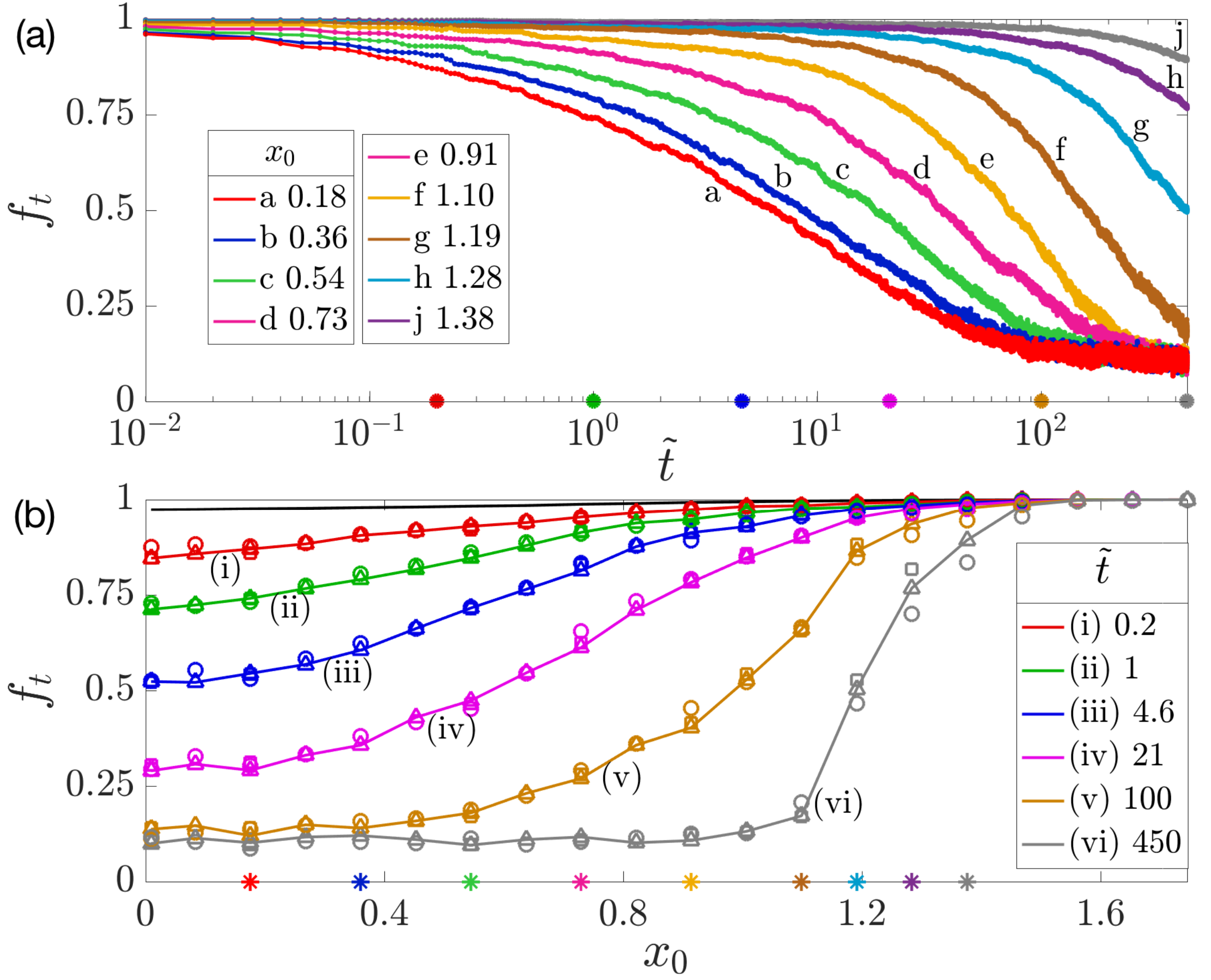}  
\caption{\label{f:dynamicsClassical}  
(color online) Classical ergodicity dynamics. For all simulations $E_0=0.74$.
\textbf{Panel (a)}: The ergodicity measure of \Eq{eq:ergodicity} as a function of time, evaluated for semiclassical simulations with selected initial $x_0$; see legend. 
\textbf{Panel (b)}: The ergodicity measure as a function of $x_0$ 
at selected moments of time; see legend.
The selected values of $\tilde{t}$ and of $x_0$ 
are indicated by filled circles and stars  
on the horizontal axis of panels~(a) and~(b), respectively.  
In panel~(b) the different symbols correspond to $v=0.2$ ($\circ$), $v=0.1$ ($\triangle$), and $v=0.05$ ($\square$). The simulations in  panel~(a) are for $v=0.1$, but look the same for other values of $v$ provided the horizontal axis is the scaled time $\tilde{t} = v^2t$.
}
\end{figure}
%%% End Figure %%%%%%%%%%%%%%%%%%%%%%%%%

%%%%%%%%%%%%%%%%%%%%%%%%%%%%%%%%%%%%%%%%%%%%%%%%%%%%%%%%%%%%%%%%%%%%%%%%%%%%%%%%%%%%%%%%%%%%%%
\section{Ergodicity measures}
\label{measures}

%%%%%%%%%%%%%%%%%%%%%%%%%%%%%%%%%%%%%%%%%%%%%%%%%%%%
\subsection{Microcanonical ergodicity measure}

Attempting to quantify ergodicity, the first inclination is to define a measure that would compare $P_{\infty}(x)$ to $P_{\text{erg}}(x)$. For that purpose we choose the overlap function
\begin{align}\nonumber
\Delta_t(x)=P_t(x)-P_\text{erg}(x)\\
\label{eq:ergodicity}
f_t=\sum_{\Delta>0} \Delta_t(x) \ \ \ \in[0,1] 
\end{align}
A limiting value of $f_{\infty}\rightarrow 0$ indicates an ergodic simulation, whereas $f_\infty\rightarrow 1$ is attained when the two distributions have almost no overlap, implying localization. 
Though very simple, the above ergodicity measure gives a reasonably good estimate that reflects the observed localization in \Fig{f:distributions}. For a smooth distribution, as is the case for   semiclassical simulations with large clouds, or quantum simulations with a large $N$, this measure becomes independent of the bin size $\delta x$.

Results for semiclassical simulations are presented in  \Fig{f:dynamicsClassical}. In panel~(a) we plot $f_t$ against the scaled time ${\tilde{t}=v^2t}$ for $E_0=0.74$ and for different initial conditions $x_0$. The dynamics display an initial transient followed by a rapid decay. A steady state is reached by some of the simulations already for $\tilde{t} \sim 100$. At $\tilde{t}\sim 450$ all simulations with ${x_0<1}$ have reached the same value $f_t \approx 0.1$. The transient time becomes exponentially large for ${x_0>1}$.   
%
%
%whose duration scales approximately as $\log (tv^2) \propto x_0$),   
% When $x_0$ corresponds to the phase space region where one of the trimers is chaotic ($|x_0|<1.23$), the ergodization rate grows slowly with $x_0$. Outside the chaotic region the ergodization rate slows down significantly.  
%
%
In panel~(b) we show $f_t$ at selected time instances for the full range of $x_0$ values. This plot demonstrates clearly that there is some kind of ``mobility edge" at ${x_0 \sim 1}$, beyond which the system does not ergodize within a physically reasonable timescale. The different symbols establish the quasistochastic scaling with respect to the strength of the coupling.    

\begin{figure*}[th!]
\centering 
\includegraphics[width=0.85\linewidth]{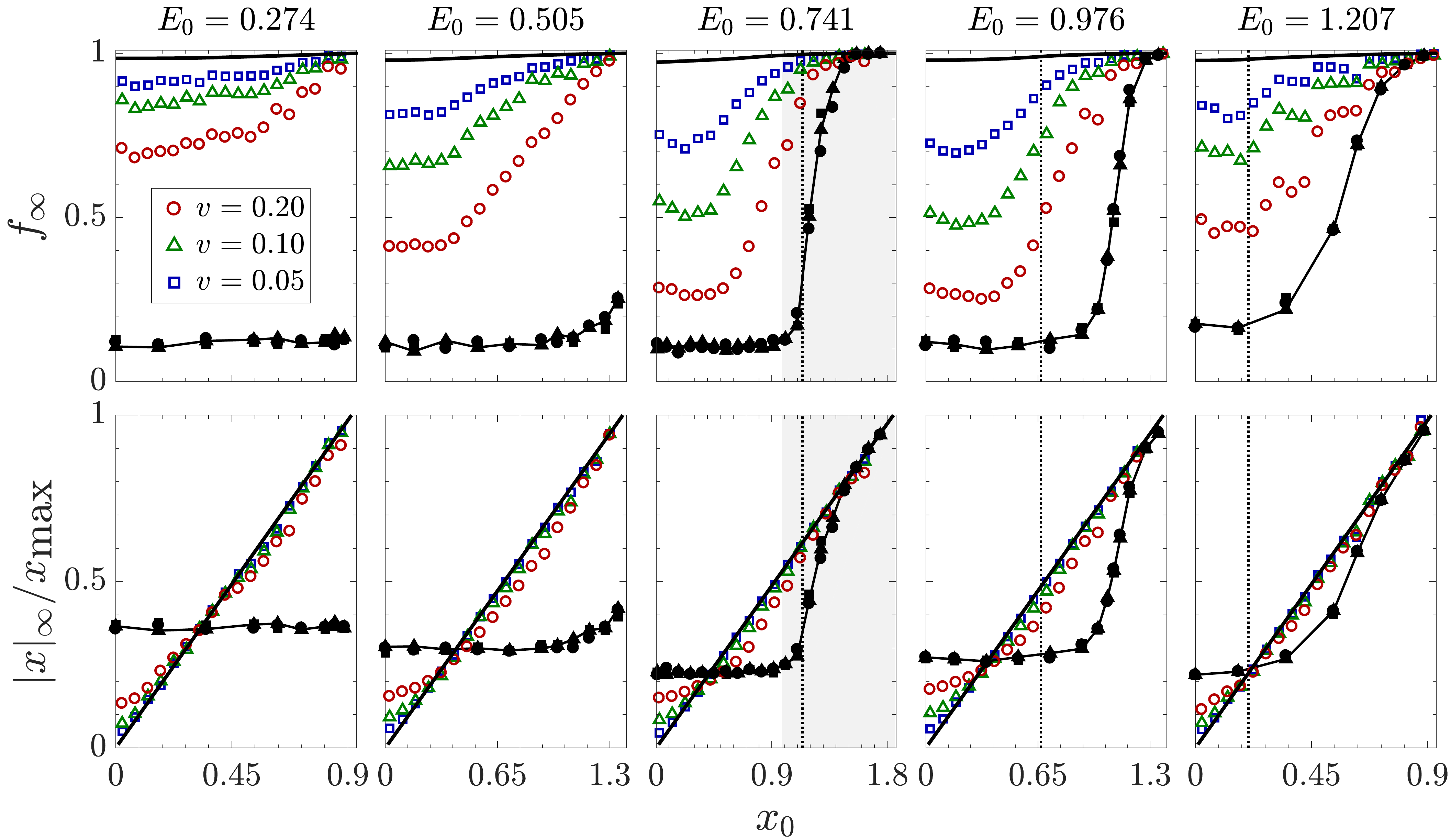}  
\caption{\label{f:localizationSummary}  
(Color online) Ergodicity and localization measures for representative simulations. Each column of panels corresponds to simulations with energy $E_0$ that is indicated at the top. Quantum results (empty markers) for system with $N_L=N_R=24$ particles are based on saturation profiles that were locally averaged over states within $E=E_0\pm 0.02$ and $x=x_0\pm 0.025$. Semiclassical results (filled markers) are evaluated for $\tilde{t}=450$, and are connected by a line. The initial conditions of the simulations are indicated by markers in \Fig{f:doubleTrimer}. Wherever applicable, the vertical lines mark the self-trapping energy, while the gray areas indicate that both trimers are quasi-integrable.
\textbf{Top panels:} The microcanonical ergodicity measure $f_{\infty}$ for different values of $v$ (see legend). The thick solid line at $f\approx 1$ indicates the value $f_{t=0}$. 
\textbf{Bottom panels:} The intrinsic ergodicity measure, namely, the normalized mean value of $|x|$, for the same simulations. See text for details. The diagonal thick line $|x|_{\infty}=x_0$ is expected to be followed if ergodicity is prevented by a strong-localization effect. 
}  
\end{figure*}

\begin{figure}[h]\centering 
\includegraphics[width=1\linewidth]{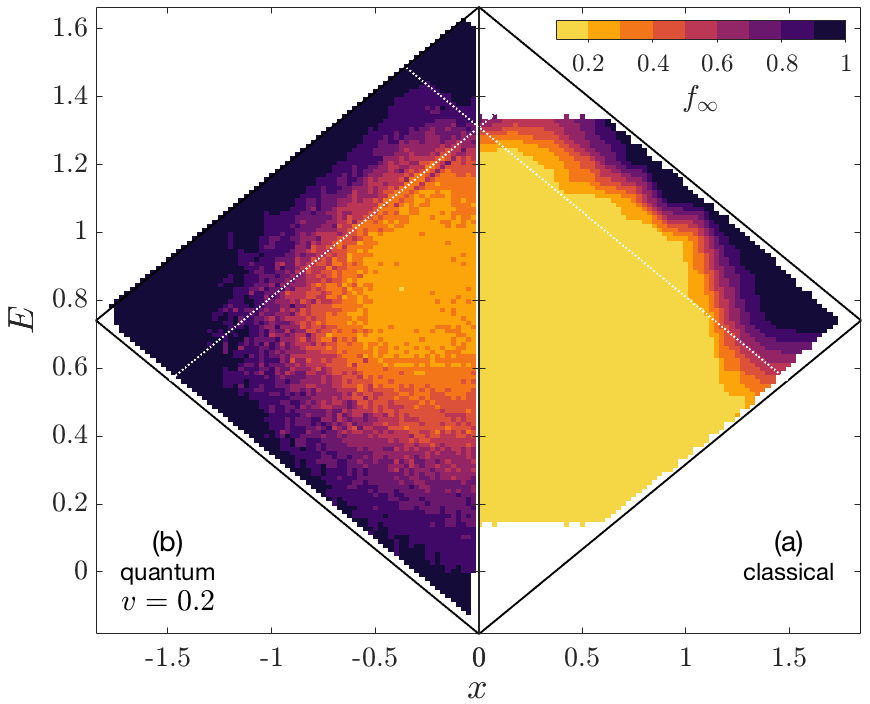}
\caption{\label{f:localizationSpectrum}  
(Color online) 
Overview of the double-trimer spectrum based on simulations as in \Fig{f:localizationSummary}. The color reflects the ergodicity measure $f_{\infty}$. Low values correspond to ergodicity, while high values imply localization. \textbf{On the right side} semiclassical results are presented (they are independent of~$v$). \textbf{On the left side} quantum results are presented for $v=0.1$. The $v$ dependence of the quantum results has been demonstrated in \Fig{f:localizationSummary} and has merely a quantitative effect. 
Semiclassical simulations in regions of low chaoticity (top and bottom corners) are missing, because they require intolerable integration accuracy. 
%The markers are the same as in \Fig{f:doubleTrimer}.
}\end{figure}

Evaluation of the ergodicity measure $f_\infty$ for semiclassical simulations with different initial energies is presented in the top row of \Fig{f:localizationSummary}. These results are also incorporated into \Fig{f:localizationSpectrum}. We see that almost all simulations, other than those lying initially within (or close to) the self-trapped region, reach $f_\infty\approx 0.1$, indicating a near-complete ergodization.

%%% Figure %%%%%%%%%%%%%%%%%%%%%%%%%%%%%
%%% End Figure %%%%%%%%%%%%%%%%%%%%%%%%%

%%% Figure %%%%%%%%%%%%%%%%%%%%%%%%%%%%%
%%% End Figure %%%%%%%%%%%%%%%%%%%%%%%%%
%%%%%%%%%%%%%%%%%%%%%%%%%%%%%%%%%%%%%%%%

%%%%%%%%%%%%%%%%%%%%%%%%%%%%%%%%%%%%%%%%%%%%%%%%%%%
\subsection{An intrinsic ergodicity measure}

It should be kept in mind that in a mixed phase space system, containing both chaotic and quasi-integrable regions, the energy surface as a whole is not ergodic. It is only the connected ``chaotic sea" region that can be regarded as ergodic. Furthermore, in the quantum case some peripheral regions of the sea may not be accessible, if the wave packet has to percolate via sub-Planck corridors. The naive comparison of $P_\infty(x)$  and $P_{\rm erg}(x)$ is hence insufficient. Instead, we would like to adopt an {\em intrinsic} definition of ergodicity that does not depend on our prejudice regarding the phase space structure, and does not involve $P_{\text{erg}}(x)$. Accordingly, 
below we use the term {\em ergodic region} for the region in which there is no memory of the initial conditions. It is the region where $P_{\infty}(x)$ becomes independent of $x_0$. By contrast, in non-ergodic regions the obtained $P_{\infty}(x)$ depends on $x_0$, with the particular extreme case of {\em localization}, where $P_{\infty}(x)$ is peaked around the initial value $x=x_0$. 

In principle, a full comparison of the distributions $P_{\infty}(x)$ for different $x_0$ is required for our memory-based intrinsic ergodicity measure. In practice, for the purpose of quantitative analysis it is more effective  to consider just a single moment of the long-time distribution, and examine its sensitivity to $x_0$. Due to the system's mirror symmetry, and since we set ${N_R=N_L}$, we choose for this purpose the function 
\beq
\avg{|x|}_\infty \ \ \equiv \ \ \sum_x |x| \, P_\infty(x)  
\eeq
For plotting purposes this measure is normalized 
by $x_{\text{max}}$, which is the highest classically allowed 
value of~$x$ for the given energy $E_0$.
The results for semiclassical simulations are presented in the bottom row of \Fig{f:localizationSummary}, and we see that they are well correlated with the microcanonical $f$~measure. 
Before considering the quantum results, which are also displayed in \Fig{f:localizationSummary}, we discuss the issue of classical localization.

%%%%%%%%%%%%%%%%%%%%%%%%%%%%%%%%%%%%%%%%%%%%%%%%%%%%%%%%%%%%%%%%%%%%%%%%%%%%%%%%%%%%%%%%%%%%%%
\section{Classical localization}
\label{classical}

The ergodic phase space region in \Fig{f:localizationSpectrum} consists of the $(E,x)$ domains where  $f_{\infty}$ is low. Outside of this region we distinguish between several classical localization mechanisms (see Appendix~A).   

The first possibility is quasi-integrability due to {\em quasilinearity}. A single trimer is not chaotic in the linear, low energy region of its spectrum. When two such trimers are coupled (the lower corner of \Fig{f:localizationSpectrum}), if $v$ is small enough the dynamics is likely to remain quasi-integrable, and therefore thermalization would be avoided. In fact, we have to restrict all such statements,  and point out that due to Arnold diffusion we can always have thermalization after an exponentially long time.

% \gmrk{[duplicate with the second paragraph in Sec.IIa]} 
The second possibility is {\em self-trapping} which is localization within an isolated part of the energy surface that is surrounded by a (higher or lower energy) forbidden region. A single trimer becomes self-trapped when its energy is above a well-defined border and is therefore dominated by interaction. 
When two self-trapped trimers are coupled (the upper corner of \Fig{f:localizationSpectrum}), or when a self-trapped trimer is coupled to a quasilinear one (the left and right corners of \Fig{f:localizationSpectrum}), energy exchange is suppressed due to the regular motion in each trimer.

By contrast, if a self-trapped trimer is coupled to a chaotic one, energy exchange should occur. This is formally like coupling an integrable oscillator to a chaotic environment for which thermalization is expected (the situation is similar to that of a harmonic oscillator under stochastic driving, or, in reverse, a system of chaotic oscillators under harmonic driving). However, in our coupled trimer system it looks like we still have classical localization (the upper right strip of \Fig{f:localizationSpectrum}). We claim that in this region the ergodic time is finite, yet has become exponentially large; an indication for that was already given in \Fig{f:dynamicsClassical}.  Below we give an expanded explanation.

A quantitative way to estimate the rate of thermalization is to use the the Fokker-Plank picture of \cite{trmPRL}. The diffusion coefficient $D$ for spreading in $x$ (within a narrow energy shell centered at $E$) is given by the classical version of the Fermi-golden-rule (see Appendix~C):
\begin{flalign}\label{eq:coefficient}
D_E(x)=\frac{v^2}{8} \sum_{j,k=1}^3 \int_{-\infty}^{\infty}\!\! \omega^2 S_{jk}(\omega,\varepsilon_L)  S_{kj}(\omega,\varepsilon_R)\frac{d\omega}{2\pi} ~,
\end{flalign}
where $\varepsilon_{L,R}=E\pm x/2$. This expression assumes that the two subsystems are independent and weakly coupled, and at least one of them has to be chaotic. 
%In the present context $S_{jj}(\omega;\varepsilon)$ is the power spectrum of the fluctuating variable $I_j(t)$ (see Appendix~C). 
The spectral density $S_{jk}(\omega;\varepsilon)$ is the Fourier transform of the cross-correlation function $C_{jk}(\tau) = \braket{I_j(t+\tau)I_k(t)}$.  
This linear-response expression implies that the dynamics for different $v$ values should be quasistochastic, meaning that the spreading is the same irrespective of $v$ when plotted against the scaled time $\tilde{t}= v^2t$.

\begin{figure}[t!]
\centering 
\includegraphics[width=1\linewidth]{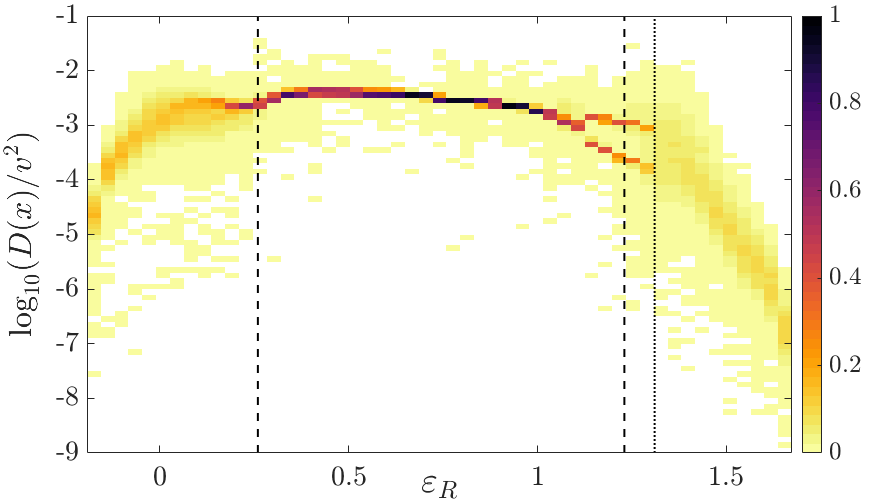}  
\caption{\label{f:diffusionCoefficient}  
The diffusion coefficient $D(x)$. Each column is a color-coded probability distribution (histogram) of $D$ values, calculated using \Eq{eq:coefficient}. The left trimer is chaotic ($\varepsilon_L=0.48$) while the energy of the right trimer ($\varepsilon_R$) is a free variable. The distribution becomes wide for a mixed-phase space. Exponentially small values are observed if the "R" trimer is self-trapped ($\varepsilon_R>1.31$). Vertical lines mark the edge of the chaotic region (dashed) and the self-trapping boundary (dotted). 
}
\end{figure}

The results for $D(x)$ are provided in \Fig{f:diffusionCoefficient}.  
A probability distribution of $D$ values (generated by using different phase space trajectories with $v=0$; see Appendix C) is presented as a function of $\varepsilon_R$, while $\varepsilon_L$ belongs to the chaotic range. Whenever the "R" trimer is self-trapped ($\varepsilon_R>1.31$), the range of frequencies associated with the motion becomes narrow, as in \Fig{f:trimer}(c). This leads to an exponentially small result. Also for small $\varepsilon_R$ the diffusion coefficient becomes small, but not exponentially small.   

Summarizing, thermalization in the classical context is generally expected whenever either one or both subsystems are chaotic. In our previous work on the trimer-monomer system, ergodization is obtained whenever the trimer is chaotic despite the integrability of the monomer. However, here we have encountered a situation where the expected thermalization process is suppressed due to a vanishingly small value of~$D$.
We note that having quasistochastic spreading is a weaker claim than having diffusion: in regions of mixed phase space the quasistochastic spreading is likely to be anomalous. Therefore we prefer to regard~$D$ in the present context as a measure of the coupling between the subsystems and not as a transport coefficient.

%%%%%%%%%%%%%%%%%%%%%%%%%%%%%%%%%%%%%%%%%%%%%%%%%%%%%%%%%%%%%%%%%%%%%%%%%%%%%%%%%%%%%%%%%%%%%%%%%%%%%%%%%%%%%%%%%%%%%%
%\clearpage

%%%%%%%%%%%%%%%%%%%%%%%%%%%%%%%%%%%%%%%%%%%%%%%%%%%%%%%%%%%%%%%%%%%%%%%%%%%%%%%%%%%%%%%%%%%%%%%%%%%%%%%%%%%%%%%%
\section{Quantum localization}
\label{quantum}

We apply the same analysis as in the classical case for the quantum saturation distributions $P_\text{sat}(x)$, which are obtained from \Eq{eq:sat_profile} with different initial values $x_0$. The results are presented in \Fig{f:localizationSummary}, and incorporated into \Fig{f:localizationSpectrum}. 
In the later figure, each pixel is color-coded according to the value of its ergodicity measure $f_\infty$. 
On the basis of these results we make several observations:
{\bf (i)}~Unlike in the classical case, there is no $x$ region where the quantum spreading is strictly ergodic. The $f$ measure is everywhere significantly larger than zero.  
{\bf (ii)}~In the quantum case there is a strong $v$ dependence. The classical limit is approached as $v$ is increased, but even for a large $v$ ergodicity is not attained. 
{\bf (iii)}~Even in the central, low-$x_0$ region we observe some memory of initial conditions. For example, focusing on $E_0=0.74$, only for the largest $v$ do we observe a tiny region (${|x_0|< 0.25}$) where the saturation is $x_0$-independent.
{\bf (iv)}~Contrary to the classical case, ergodization is strongly suppressed for large $|x_0|$ even in the low-energy regime (small $E_0$). 

There is a subtle point that should be emphasized. In the classical case, there is a vast $x_0$ region where the numerical simulations yield ergodic distributions, as indicated by the low $f$ measure and by the $x_0$-independence of $\braket{|x|}_{\infty}$. However, we cannot rule out the possibility of a slow classical ergodization over exponentially long timescales. Thus, it is possible that the observed picture of classical localization is only a transient. The same reservation does not apply in the quantum case, since the saturation profile after an infinitely long time can be obtained via direct diagonalization using \Eq{eq:sat_profile}.

%%%%%%%%%%%%%%%%%%%%%%%%%%%%%%%%%%%%%%%%
%%% Figure %%%%%%%%%%%%%%%%%%%%%%%%%%%%%
%%% End Figure %%%%%%%%%%%%%%%%%%%%%%%%%
%%%%%%%%%%%%%%%%%%%%%%%%%%%%%%%%%%%%%%%%

%%%%%%%%%%%%%%%%%%%%%%%%%%%%%%%%%%%%%%%%%%%%%%%%%%%%%
\subsection{Perturbative localization}

When the hopping between two unperturbed states of the decoupled system is small compared with their energy difference, elementary perturbation theory stipulates that the true eigenstates of the coupled system remain localized (no mixing). For time-dependent dynamics, this means that the system cannot make a transition between the unperturbed eigenstates. This type of {\em perturbative} localization is trivial, and we would like to verify that the localization we observe is more profound. For this purpose, we calculate and display in \Fig{f:saturation}(a) the $x$-basis participation number: 
\beq\label{eq:M}
M \equiv \Big[ \sum_x P(x)^2 \Big]^{-1}~.
\eeq
Whenever the localization is due to perturbative localization we expect to observe $M\sim 2$, with $M=2$ being the minimal value for $x_0 \neq 0$ due to the mirror symmetry. This {\em perturbative localization} takes place whenever the $v$-related couplings are small compared to the mean level spacing. In our system the density of states becomes smaller at peripheral (large $|x|$) regions; hence localization there might be of this trivial type.

%%%%%%%%%%%%%%%%%%%%%%%%%%%%%%%%%%%%%%%%%%%%%%%%%%%%%
\subsection{Localization due to break time}

\begin{figure}
\centering 
\includegraphics[width=1\linewidth]{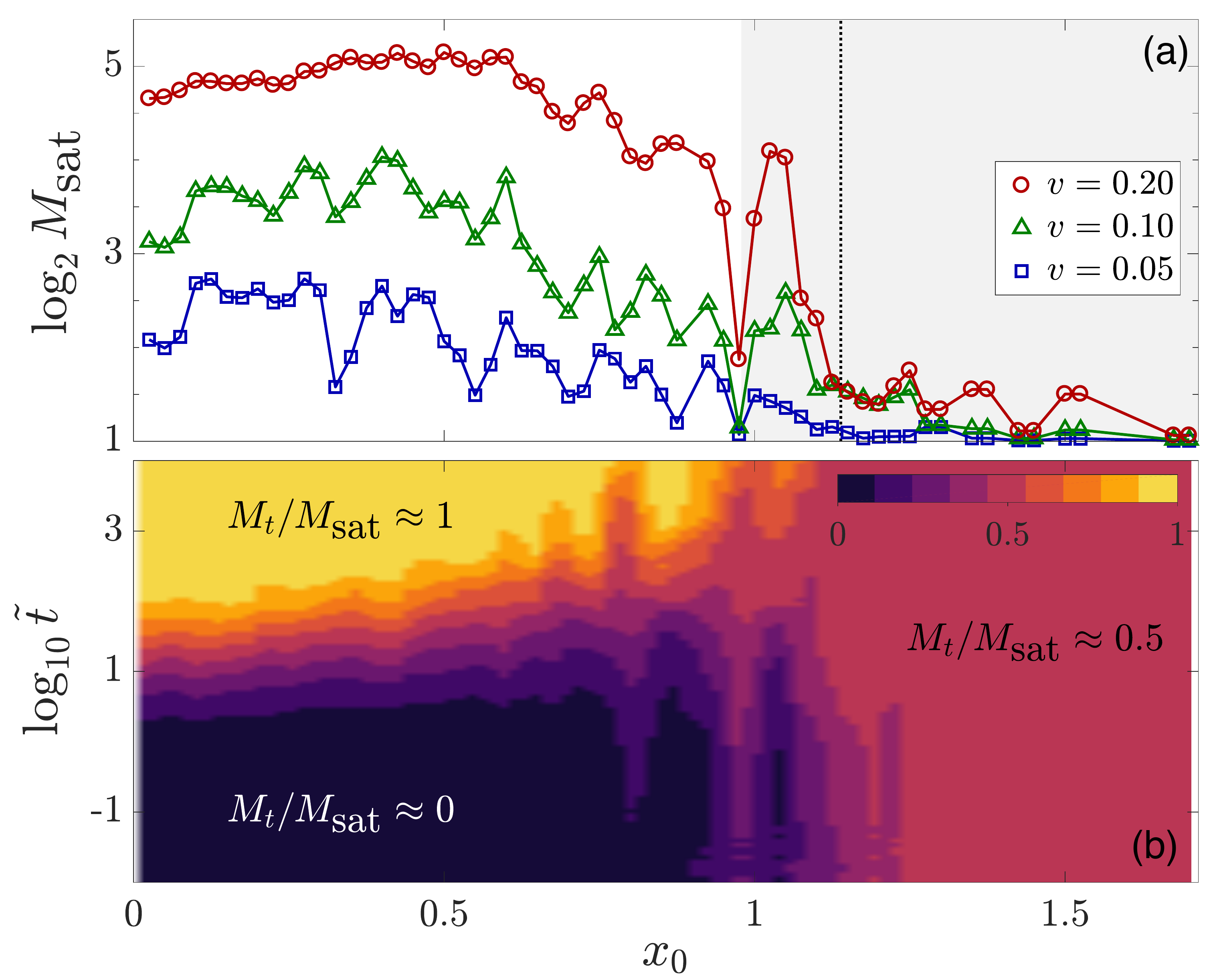}  
\caption{\label{f:saturation} 
(Color online) Saturation of the quantum spreading. The parameters are the same as in the $E_0=0.741$ column of \Fig{f:localizationSummary}.  
\textbf{Panel (a)} The participation number for the quantum saturation profile, locally averaged over states within $E=E_0\pm 0.02$ and $x=x_0\pm 0.025$. The vertical line marks the self-trapping energy. In the gray area both trimers are quasi-integrable.
\textbf{Panel (b)} The approach to saturation. The ratio  $M_t/M_{\text{sat}}$ is imaged as a function of time (here ${v=0.2}$). In the quasi-ergodic regime long times are required in order to resolve the  quasidegeneracies, and to get unit ratio.  
}
\end{figure}

The saturation values of $M$, displayed in \Fig{f:saturation}, are calculated for the same simulations as in the central, $E_0=0.741$ column of \Fig{f:localizationSummary}. In regions where $M_\text{sat}$ is significantly larger than unity, simple semiclassical reasoning suggests ergodization. Thus, naively, localization is not expected there. A more sophisticated semiclassical reasoning does suggest the {\em possibility} of a dynamical localization effect due to a break time in the quantum-to-classical correspondence. We have refined and tested this approach for the trimer-monomer system \cite{Chris}, where we have demonstrated that the quantum break time (and hence the regions where localization shows up) can be deduced from careful semiclassical analysis. In  the present study, we do not aim to repeat the same type of analysis for the double-trimer configuration,  but rather to determine whether nontrivial dynamical localization shows up.  

A general argument that supports our expectation to observe nontrivial dynamical localization suggests the existence of a {\em mobility edge} based on a semiclassical picture. According to the break-time phenomenology, the quantum dynamics fails in following the classical spreading if the classical exploration of phase space becomes sub-ballistic. This is always the case for diffusion in one dimension, and also (marginally) always the case in two dimensions, but not in higher dimensions. Still, in higher dimensions we can expect the formation of a mobility edge if the ballistic exploration becomes slow enough. (Note that the term ``exploration" is not a synonym for ``spreading". Standard diffusive spreading of a cloud in high dimensions involves a ballistic exploration of phase space cells.) We argue that a slow rate of exploration is typical for systems with mixed phase space. At the peripheral regions of the chaotic sea the timescales for migration between different regions become very slow. This is indeed verified in the present example, where we witness, as $|x_0|$ approaches the phase space boundary, a very small $D$ or even exponentially small $D$. Respectively, at low energies the small $D$ arises due to quasi-integrability, while for large energies the exponentially small $D$ is due to self-trapping. Indeed in the quantum simulations, see \Fig{f:localizationSpectrum}, we witness a mobility edge in both cases. 
   
Looking again at \Fig{f:saturation}(b) we clearly see that in the quantum case there is a residual memory of $x_0$, which is a signature of a nontrivial dynamical localization effect. This lack of quantum ergodicity persists also in regions where $M$ is significant larger than~2. The region where we can identify a small quantum-ergodic domain that features $x_0$-independent saturation appears only for relatively large coupling ($v=0.2$), and it is rather small (${|x_0|< 0.25}$) as opposed to the classically observed ergodicity.

%%%%%%%%%%%%%%%%%%%%%%%%%%%%%%%%%%%%%%%%
%%% Figure %%%%%%%%%%%%%%%%%%%%%%%%%%%%%
%%% End Figure %%%%%%%%%%%%%%%%%%%%%%%%%
%%%%%%%%%%%%%%%%%%%%%%%%%%%%%%%%%%%%%%%%

%%%%%%%%%%%%%%%%%%%%%%%%%%%%%%%%%%%%%%%%%%%%%%%%%%%%%
\subsection{The approach to saturation}

Remember that the whole discussion of quantum localization does not require any simulations, since the formal saturation profile $P_{\infty}(x)$ can be obtained via \Eq{eq:sat_profile}. Optionally, we could have used \Eq{eq:P_x} with some large $t$. However, as explained below, the two procedures are not practically equivalent. 

Besides the quasistochastic timescale of the semiclassical dynamics, we have another (much longer) timescale that has to do with the quasidegeneracies of the spectrum (see Appendix~D). Due to the mirror symmetry of the double-trimer system, it is clear that all eigenstates have a definite parity and therefore the $x$-projected saturation profile satisfies $P_{\text{sat}}(-x)~=~P_{\text{sat}}(x)$. Contrary to this expectation, some long-time-averaged simulations give ${P_{\infty}}(-x)~\ne~{P_{\infty}}(x)$. This happens due to quasidegeneracies in the perturbed  spectrum that are present for small $v$ and can be barely resolved in a finite-time simulation. These quasidegeneracies are remnants of the exact $v=0$ degeneracy of the $\ket{E,\pm x}$ states.

The time evolution of $M_{\infty}/M_{\text{sat}}$ is provided in \Fig{f:saturation}(b). This ratio can be regarded as a measure for the relative contribution of quasidegeneracies. Whenever the the initial state significantly overlaps quasidegenerate pairs, we get for long time ${P_t(x) = 2P_{\text{sat}}(x)}$ and therefore $M_t/M_{\text{sat}}=1/2$. A much larger timescale is required in order to recover  ${P_{\infty}(x) = P_{\text{sat}}(x)}$.  
This discussion clarifies our choice of  $\braket{|x|}_{\infty}$ rather than $\braket{x}_{\infty}$ as a measure for the characterization of the spreading profile. Strictly, the latter will be always get to zero in the infinite time limit irrespective of localization, merely due to the mirror symmetry of the system.

%%%%%%%%%%%%%%%%%%%%%%%%%%%%%%%%%%%%%%%%%%%%%%%%%%%%%%%%%%%%%%%%%%%%%%%%%%%%%%%%%%%%%%%%%%%%%%%%%%%%%%%%%%%%%%%%
%%%%%%%%%%%%%%%%%%%%%%%%%%%%%%%%%%%%%%%%%%%%%%%%%%%%%%%%%%%%%%%%%%%%%%%%%%%%%%%%%%%%%%%%%%%%%%%%%%%%%%%%%%%%%%%%
\section{Summary}
\label{conclusion}

Localization is commonly viewed as an interference phenomenon that leads to a breakdown of quantum-classical correspondence. Such definition is not {\em intrinsic}: it requires an \textit{a priori} definition of some reference space and a measure that compares the actual quantum spreading to a different (classical) dynamics. This point of view is quite frustrating when dealing with small quantized systems, because in practice the correspondence is quite bad to begin with. We therefore prefer to view localization as a lack of ergodicity, and to provide a measure for {\em intrinsic ergodicity}. We have defined ergodicity as the case of having some space within which the saturation profile becomes independent of the initial conditions. Accordingly, we could analyze {\em separately} the ergodicity in the classical case and in the quantum case, and then independently compare the two.        

One should remember that several mechanisms can lead to {\em localization}. The most trivial one, as in the case of a single trimer with large $u$, is self-trapping. This is classical localization due to energetic stability within an isolated region of the energy surface. A more complex type of classical localization arises due to dynamical stability of quasilinear motion. The latter case is endangered by Arnold diffusion, which, in practice, can be ignored due to unrealistic time scales. 

In realistic models of the type we study, where we have mixed phase space, the microcanonical ETH assumption \cite{Olshanii} is typically violated, because vast regions of the energy shell are not accessible.  Even if the classical dynamics is ergodic on very long time scales, the quantum will fail to penetrate peripheral regions due to the dynamical or perturbative localization effect, and therefore the expectation value of the observable will not approach the microcanonical one as expected from ETH. Our definition of quantum ergodicity does not make any assumption about the ergodic distribution (microcanonical or not).

Our aim in this work was to highlight the emergence of the {\em quantum} dynamical localization effect in the thermalization process of weakly coupled subsystems. The term {\em weak coupling} is used in a classical sense; quantum mechanically, the coupling might lead to mixing of many levels. In fact, the extreme case of weak coupling in the quantum sense is not very interesting at all, because it leads to a {\em perturbative localization}, indicated by ${M \sim 2}$ participation numbers (i.e., no spreading happens). However, we do observe clear signs of quantum localization even when the coupling is not trivially small, so that $M$ is large. 
We were thus able to demonstrate a quantum mechanical loss of {\em intrinsic ergodicity} despite having nonperturbative quantum mixing.      
This is the nontrivial type of quantum dynamical localization that we were looking for.

%%%%%%%%%%%%%%%%%%%%%%%%%%%%%%%%%%%%%%%%%%%%%%%%%%%%%%%%%%%%%%%%%%%%%%%%%%%%%%%%%%%%%%%%%%%%%%%%%%%%%%%%%%%%%%%%

\ \\

\sect{Acknowledgment} 
The present line of study has been inspired by discussions with Shmuel Fishman who passed away recently (2-April-2019).
This research was supported by the Israel Science Foundation (Grant  No.~283/18).

%%%%%%%%%%%%%%%%%%%%%%%%%%%%%%%%%%%%%%%%%%%%%%%%%%%%%%%%%%%%%%%%%%%%%%%%%%%%%%%%
%%%%%%%%%%%%%%%%%%%%%%%%%%%%%%%%%%%%%%%%%%%%%%%%%%%%%%%%%%%%%%%%%%%%%%%%%%%%%%%%
%%%%%%%%%%%%%%%%%%%%%%%%%%%%%%%%%%%%%%%%%%%%%%%%%%%%%%%%%%%%%%%%%%%%%%%%%%%%%%%%
\newpage
\appendix

%%%%%%%%%%%%%%%%%%%%%%%%%%%%%%%%%%%%%%%%%%%%%%%%%%%%%%%%%%%%%%%%%%%%%%%%%%
\section{The eigenstates of a single trimer}

The Hamiltonian of a one-particle linear trimer (${N_\alpha=1}$ and ${u=0}$) is diagonal in the basis
\beq
\begin{aligned}
&\ket{\text{sym}}=\frac{1}{2}(\ket{1}+\sqrt{2}\ket{2}+\ket{3})~,\\
&\ket{\text{anti}}=\frac{1}{2}(\ket{1}-\sqrt{2}\ket{2}+\ket{3})~,\\
&\ket{\text{dark}}=\frac{1}{\sqrt{2}}(\ket{1}-\ket{3})~.\\
\end{aligned}
\eeq
with the corresponding energies
\beq
\omega_\text{sym}=-\omega_0~,\ \ \ 
\omega_\text{anti}=+\omega_0~,\ \ \ 
\omega_\text{dark}=0~,
\eeq
where ${\omega_0 = K/\sqrt{2}}$.
We can define operators which annihilate a particle in one of these orbitals,
\beq
\begin{aligned}
&\hat{a}_\text{sym}\ =\ \frac{1}{2}(\hat{a}_1+\sqrt{2}\hat{a}_2+\hat{a}_3)~,\\
&\hat{a}_\text{anti}\ =\ \frac{1}{2}(\hat{a}_1-\sqrt{2}\hat{a}_2+\hat{a}_3)~,\\
&\hat{a}_\text{dark}\ =\ \frac{1}{\sqrt{2}}(\hat{a}_1-\hat{a}_3)~,\\
\end{aligned}
\eeq
Using the corresponding occupation operators ${n=a^\dagger a}$ the many-body Hamiltonian can be rewritten as
\beq \label{eH0}
H_\text{trimer}\ =\ \omega_0(\hat{n}_\text{anti} - \hat{n}_\text{sym})~.
\eeq
The classical values ${n_j/N_\alpha}$, or the quantum expectation values ${\avg{n_j}/N_\alpha}$, are denoted by $I_j$.

If the spectrum of the trimer Hamiltonian was presented on a 2D grid ${(\varepsilon,I_\text{dark})}$ it would form a triangular lattice of points representing all possible occupations, with corners at ${(-\omega_0,0)}$, ${(0,1)}$, and ${(\omega_0,0)}$.  
These corners correspond to coherent states, meaning that all particles occupy a single orbital. More generally, each eigenstate can be characterized by its one-particle purity. For this purpose we define the one-particle probability matrix      
\beq
\rho^{(1)}_{jk} \ = \ \frac{1}{N}\avg{\hat{a}_j^\dagger \hat{a}_k}~.
\eeq
The purity is then defined as
\beq
\gamma = \text{purity} \ = \ \text{trace}\left\{[\rho^{(1)}]^2\right\} \ \in \ [1/3,\, 1]~.
\eeq
The inverse purity tells us what is the number of one-particle states (whether orbitals or site states) that ``participate" in the formation of the many-body state, ranging from 1 (e.g., for coherent states) to 3 (maximally mixed states).

\begin{figure*}[th!]
\centering 
\includegraphics[width=0.9\linewidth]{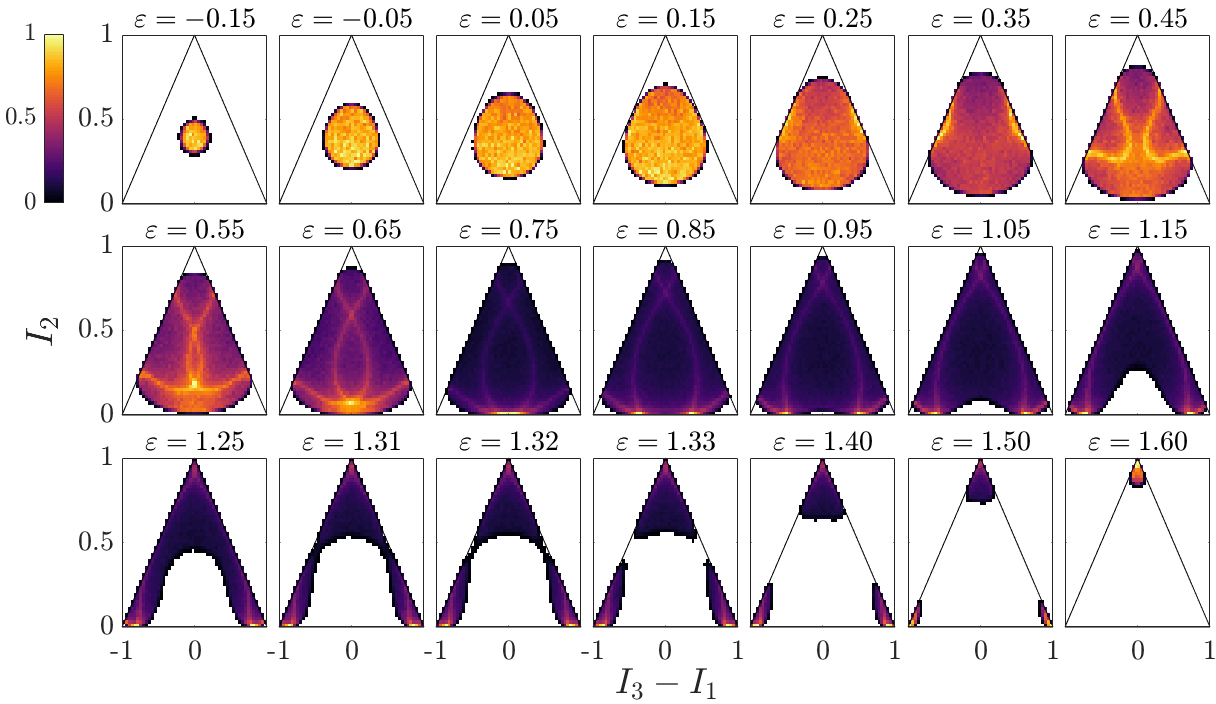}  
\caption{\label{f:trimerDOS}  
(Color online) Energy surfaces of the trimer projected onto $I$ space. Each surface is numerically represented by a uniform microcanonical cloud of classical points. The color reflects the projected (normalized) density of the points, and hence provides the (normalized) local density of states. The interaction parameter is $u=3$. The entire spectrum lies within the range $(-0.18,1.66)$. Self-trapping is observed above $\varepsilon=1.31$.
}
\end{figure*}

Once the interaction is turned on (${u>0}$), the situation changes. The energy landscape is illustrated in \Fig{f:trimerDOS}. At the low energy range one has remnants of (almost) unperturbed eigenstates (purity~$\approx1$). In the upper energy range one should distinguish between two borders whose determination is discussed in Appendix~B. The first one is the edge of the chaotic sea: up to this energy each eigenstate resembles a random wave that dwells on the classically chaotic part of the energy surface (purity~$\approx1/3$).
The second border is the threshold for self-trapping: above this value the energy surface is composed of 3 disjoint regions around $I_j \sim 1$ with ${j=1,2,3}$, with classical clouds being unable to migrate between these regions. Quantum mechanically, we see the formation of states belonging to two distinct types: states that predominantly occupy the middle site (purity~$\approx1$), or cat superpositions of states that are self-trapped in either site 1 or site 3 (purity~$\approx1/2$).

%%%%%%%%%%%%%%%%%%%%%%%%%%%%%%%%%%%%%%%%
%%% Figure %%%%%%%%%%%%%%%%%%%%%%%%%%%%%
%%% End Figure %%%%%%%%%%%%%%%%%%%%%%%%%
%%%%%%%%%%%%%%%%%%%%%%%%%%%%%%%%%%%%%%%%

%%%%%%%%%%%%%%%%%%%%%%%%%%%%%%%%%%%%%%%%%%%%%%%%%%%%%%%%%%%%%%%%%%%%%%%%%%%%%%%%%%%%%%%%%%%%%%
\section{Chaotic range determination}

Chaotic classical motion can be identified by directly observing the dynamics of individual trajectories on various energy surfaces, for example, via plotting Poincar\'e sections. 
Another possibility is to inspect the power spectrum for a variable of interest: a defining property of a chaotic motion is that the Fourier transform of a fluctuating variable exhibits a continuous frequency spectrum. By contrast, the Fourier spectrum of a regular trajectory is made of zero-width delta peaks. 

Alternatively, the underlying classical chaos is indicted in the spectral statistics of the many-body system. Let $\varepsilon_n$ be an ordered set of energy levels, and ${s_n\equiv\varepsilon_{n+1}-\varepsilon_n}$ the nearest-neighbor level spacings. Following Ref.~
\cite{r_measure} we consider the ratios   
\beq\label{eq:r}
r_n \ = \ \frac{\text{min}\left\{s_n,s_{n+1} \right\}}{\text{max}\left\{s_n,s_{n+1} \right\}} \ \ \in  \ \ [0,1]
\eeq
For classically integrable systems with two or more degrees of freedom, the quantum levels typically tend to cluster. If a parameter in the Hamiltonian is varied, level crossing may be observed. The levels appear to arise from uncorrelated events in a Poissonian random process, resulting in the probability distribution ${P(r)=2/(1+r^2)}$ \cite{r_measure}. Its average value ${\avg{r}_\text{Poisson}\approx0.386}$ is therefore considered representative for integrable classical motion.

In the opposite case, for classically nonintegrable systems whose phase space is dominated by chaos, the levels are correlated and a strong repulsion is observed. For such systems the spectral statistics are typically modeled by random matrix theory methods \cite{HaakeTextbook}. In particular, the Gaussian orthogonal ensemble (GOE) is employed for time-reversal-invariant systems. A numerical estimation of the distribution of $r$ yields ${\avg{r}_\text{GOE}\approx0.530}$ \cite{r_measure}; this value is indicative of fully chaotic classical motion. 

A practical determination of the chaotic energy range is achieved by locating the energies at which the distribution of $r$ values agrees to an equal degree with both limiting cases. 
In \Fig{f:trimer}(b) we plot the local average $\avg{r}$, taken over many levels within a small energy window, based on the spectrum of the trimer Hamiltonian, \Eq{eq:Hamiltonian_alpha}. Chaos is predominant for ${0.26<\varepsilon<1.23}$, since in this region of the spectrum we have ${\avg{r}>(\avg{r}_\text{Poisson}+\avg{r}_\text{GOE})/2}$. 

\sect{Remarks}
One should be careful when considering energies lying in the parts of the spectrum where the density of states is low. The number of levels should be sufficient for meaningful statistics, otherwise the local average $\avg{r}$ becomes strongly dependent on the size of the averaging window. In \Fig{f:trimer}(b) those bad regions were removed. 

Additionally, when evaluating the distribution of $r$ values one must take into account only levels with the same values for all conserved observables except the Hamiltonian. For the trimer this implies using only states of same parity (see Appendix~D).

%%%%%%%%%%%%%%%%%%%%
\sect{The trimer spectrum} 
A detailed examination of \Fig{f:trimer}(b) shows that fully chaotic classical dynamics is expected in the range ${0.45<\varepsilon<1}$, where ${\avg{r}\approx \avg{r}_\text{GOE}}$. Indeed, as verified by Poincar\'e sections, a typical phase space trajectory explores the entire energy surface. For other energies, where $\avg{r}$ is lower, the phase space is mixed; here regular ``islands" exist within a chaotic ``sea". In both the low- and high-energy edges of the spectrum we have ${\avg{r}\approx \avg{r}_\text{Poisson}}$, indicating quasi-integrability, as explained below.

%%%%%%%%%%%%%%%%%%%%
\sect{The trimer landscape} 
The trimer energy landscape is presented in \Fig{f:trimerDOS}. 
Close to the ground state, ${I_1\approx I_2\approx I_3}$ everywhere on the energy surface. The classical Hamiltonian, \Eq{eq:Energy_alpha}, is predominated by its linear term; without a sufficiently strong nonlinearity chaos cannot develop. 
At high energies the situation is opposite: here the nonlinearity is too strong, causing the phase space to fragment into isolated regions above ${\varepsilon=1.31}$ (self-trapping). A phase space trajectory remains forever trapped in its initial region, and hence can never explore the entire energy surface. This reduced exploration means that chaos cannot develop, and hence quasi-integrability is restored. [In fact, a weaker form of self-trapping appears already for ${\varepsilon>1}$, when some trajectories become bounded to one side of the line ${I_1=I_3}$ for very long times. This is the cause of the sudden increase in the dispersion of $D(x)$ in \Fig{f:diffusionCoefficient}.]

%%%%%%%%%%%%%%%%%%%%%%%%%%%%%%%%%%%%%%%%%%%%%%
\section{Classical simulations}

%%%%%%%%%%%%%%%%%%%%%%%%%%%%%%%%%%%%%%
\sect{Constructing a semiclassical cloud}
The initial condition for any semiclassical simulation is a cloud of 1000 points in the phase space of the double trimer, uniformly distributed over the entire energy surface $(\varepsilon_{L0},\varepsilon_{R0})$, or equivalently, $(E_0,x_0)$. Below we outline the procedure used to generate this distribution.

Consider a single trimer and let the set ${P=\left\{I_{1,2,3},\varphi_{1,2,3}\right\}}$ represent a phase space point belonging to an energy surface $\varepsilon_0$. First, a large number of points is generated by making random draws from a uniform distribution, taking ${\varphi_{1,2,3}\in[0,2\pi)}$ and ${I_{1,2}\in[0,1]}$. The third occupation is calculated from ${I_3=1-I_1-I_2}$; all points for which ${I_3<0}$ are nonphysical and thus discarded. Next, we use \Eq{eq:Energy_alpha} to evaluate the energies $\varepsilon_P$ associated with each point. Whenever the result is too far from the required value (in practice, when ${|\varepsilon_P-\varepsilon_0|>10^{-4}}$), that point is discarded. 
Finally, after collecting 1000 good points for each trimer, we pair them at random (taking one from each trimer) and evaluate the complex amplitudes ${b_\aj=\sqrt{I_\aj}\eexp{i\varphi_\aj}}$. Dynamics of the entire semiclassical cloud are generated by independently propagating each point under the equations of motion ${d b_\aj/dt=-i\partial \mathcal{H}^{cl}/\partial b_\aj^*}$, where $\mathcal{H}^{cl}$ is given by \Eq{eq:Energy_total}.

%%%%%%%%%%%%%%%%%%%%%%%%%%%%%%%%%%%%%%
\sect{Trimer power spectrum}
Consider a trimer phase space trajectory moving on an energy surface $\varepsilon$. The trajectory generates a fluctuating variable $I_j(t)$, for which we define a scaled Fourier transform 
\beq\label{eq:Y_j}
Y_j(\omega) \ \ \equiv \ \ 
\lim_{T\rightarrow \infty}\frac{1}{\sqrt{T}}\int_0^T I_j(t) e^{-i\omega t} dt~.
\eeq
The fluctuations of $I_j(t)$ are quantified by its power spectrum,
\beq
S_{jj}(\omega;\varepsilon) \ \ = \ \ |Y_j(\omega)|^2~.
\eeq
When the phase space is fully chaotic, each trajectory eventually explores the entire energy surface, and therefore has the same frequency content. However, a numerical estimate of $Y_j(\omega)$ is always limited by the total simulation time $T$ and the resolution $dt$; to reduce inaccuracies in $S_{jj}(\omega;\varepsilon)$, one can average over the spectra generated by several trajectories. Furthermore, in the case of a mixed phase space each trajectory may have a different frequency content, depending on the explored phase space regions. A microcanonical averaging (i.e., over a cloud of trajectories initially uniformly spread over the energy surface) gives a smoothed power spectrum representative of the typical motion on the energy surface. 
The power spectrum of $I_2(t)$, generated by such a cloud containing 1000 trajectories for each value of $\varepsilon$, is presented in \Fig{f:trimer}(c).

More generally we define also cross-correlation functions,
\beq
C_{jk}(\tau) \ \ \equiv \ \ \langle I_{j}(t+\tau) I_{k}(t) \rangle~. 
\eeq
Their Fourier transforms are
\beq 
S_{jk}(\omega;\varepsilon) \ \ = \ \ {Y}^*_j(\omega)Y_k(\omega)~, 
\eeq

%%%%%%%%%%%%%%%%%%%%%%%%%%%%%%
\sect{Diffusion coefficient}
The classical equations of motion for the double-trimer system yield
\beq
\dot x = -\frac{v}{4} \sum_{j=1}^3 [\dot{I}_{R j}(t)I_{L j}(t) - \dot{I}_{L j}(t)I_{R j}(t) ]~.
\eeq
The spreading in $x$ can be estimated from
\begin{flalign}\label{eq:varX}\nonumber
\langle[x(t)&-x(0)]^2\rangle = \left(\frac{v}{4}\right)^2 \sum_{j,k=1}^3  \int_0^t  \int_0^t dt' dt''  \\ \nonumber
&\times \ \langle\ [\dot{I}_{R j}(t')I_{L j}(t') - \dot{I}_{L j}(t')I_{R j}(t') ]  \\
&\times \, [ \dot{I}_{R k}(t'')I_{L k}(t'') - \dot{I}_{L k}(t'')I_{R k}(t'') ]\ \rangle
\end{flalign}
where $\langle\cdots\rangle$ implies a microcanonical average. Assuming that the motion is chaotic, it makes sense to regard the $I(t)$ functions as stationary noisy functions. 
For small enough $v$ we can write the right-hand side in terms of cross-correlation functions of the "L" and "R" uncoupled trimers: 
\begin{flalign}\nonumber
& t \times \left(\frac{v}{4}\right)^2 \sum_{j,k=1}^3  \int_{-t}^t \left[2\frac{d}{d\tau}C^{[L]}_{jk}(\tau)\,\frac{d}{d\tau}C^{[R]}_{jk}(\tau) \right.\\
&\left. -C^{[R]}_{jk}(\tau)\frac{d^2}{d\tau^2}C^{[L]}_{jk}(\tau) -C^{[L]}_{jk}(\tau)\frac{d^2}{d\tau^2}C^{[R]}_{jk}(\tau) \right]d\tau  ~.
\end{flalign}
Assuming a short correlation time, and disregarding
drift (which can be always neglected for short times), we deduce that
\begin{flalign}
\text{var}(x)=\frac{v^2t}{4} \sum_{j,k=1}^3  \int_{-\infty}^\infty\!\! \omega^2 S_{jk}(\omega;\varepsilon_L)S_{kj}(\omega;\varepsilon_R)\frac{d\omega}{2\pi}~.
\end{flalign}
From the definition ${\text{var}(x)=2Dt}$  we arrive at \Eq{eq:coefficient}. 
In practice, $D$ is evaluated from finite trajectories. 
Since the trimer phase space is mixed, some of those trajectories are quasi-integrable and some are chaotic. Thus, from an ensemble of trajectories we get some distribution of $D$ values, as shown in \Fig{f:diffusionCoefficient}.

\begin{figure}
\centering 
\includegraphics[width=1\linewidth]{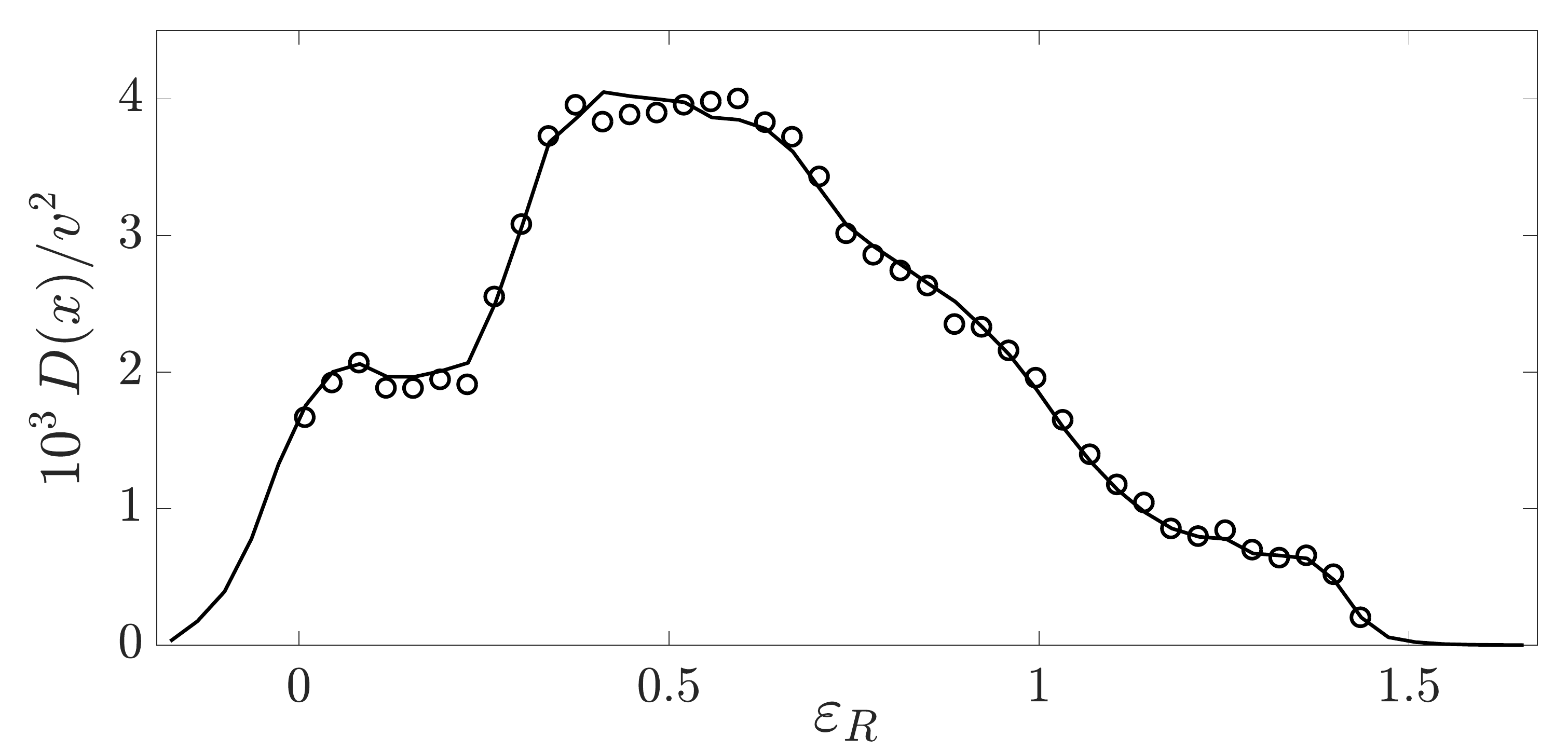}  
\caption{\label{f:diffusionVariance}  
Testing the validity of \Eq{eq:coefficient}. The line shows the mean value of the $D$ distribution in \Fig{f:diffusionCoefficient}. Symbols correspond to a linear fit ${\partial \text{var}(x)/\partial t=2D}$, using semiclassical simulations for a double trimer with $v=0.01$. In all cases $\varepsilon_L=0.48$.
}
\end{figure}

The theoretical estimate of \Eq{eq:coefficient}, where formally $v\rightarrow 0$, can be tested by using semiclassical simulations with a small $v$. We evaluate ${\partial\, \text{var}(x)/\partial t=2D}$ for the time range ${0.01<v^2t <0.1}$, where to a good approximation the variance grows linearly with time. The results are shown in \Fig{f:diffusionVariance} (symbols) as a function of $\varepsilon_R$, with $\varepsilon_L$ being kept constant. The theoretical $D$ value is given by the mean of the $D$ distribution for the same values of $\varepsilon_{L,R}$ (line). The agreement is excellent across a wide range of energies, even when the ``R" trimer is within the self-trapping region (${\varepsilon_R>1.31}$).

%%%%%%%%%%%%%%%%%%%%%%%%%%%%%%%%%%%%%%%%
%%% Figure %%%%%%%%%%%%%%%%%%%%%%%%%%%%%
%%% End Figure %%%%%%%%%%%%%%%%%%%%%%%%%
%%%%%%%%%%%%%%%%%%%%%%%%%%%%%%%%%%%%%%%%

%%%%%%%%%%%%%%%%%%%%%%%%%%%%%%%%%%%%%%%%%%%%%%
\section{Symmetry properties of eigenstates}

A single trimer, shown in the inset of \Fig{f:trimer}(a), has one nontrivial symmetry operation: the exchange of its edge sites. Hence, the eigenstates of the trimer Hamiltonian $\mathcal{H}_\alpha$ of \Eq{eq:Hamiltonian_alpha} are either symmetric or antisymmetric under this operation. The Hilbert space in which they reside is split into two subspaces, each containing only states of the same parity, with the symmetry imposing $I_1=I_3$ for all eigenstates. 

The double trimer, shown in the inset of \Fig{f:doubleTrimer}, contains three nontrivial symmetry operations: $P_1$ that exchanges  edge sites in both trimers, ${(\alpha1\leftrightarrow \alpha3,\ \alpha=L,R)}$; $P_2$ that exchanges all corresponding sites between the two trimers, ${(Lj\leftrightarrow Rj,\ j=1,2,3)}$; and their product $P_1P_2$. We label the states symmetric or antisymmetric under $P_1$ ($P_2$) by $S$ or $A$ ($s$ or $a$), respectively. 
The resulting four irreducible representations (correspondingly labeled by $Ss$, $Sa$, $As$, and $Aa$) are associated with four Hilbert subspaces. A new basis composed of superpositions of Fock states, each one symmetric or antisymmetric under $P1$ and $P2$, takes the Hamiltonian of \Eq{eq:Hamiltonian_parts} into a block-diagonal form, with approximately equal-sized blocks. This transformation significantly reduces the computational difficulty of exact diagonalization, making it possible to reach higher particle numbers.

When discussing energy exchange between weakly coupled trimers, the most intuitive energy basis is the set of direct products of the single-trimer eigenstates, $\ket{E,x}=\ket{\varepsilon_n}\ket{\varepsilon_m}$, with the mean energy $E=\varepsilon_n/2+\varepsilon_m/2$ and the imbalance $x=\varepsilon_n-\varepsilon_m$. A generic direct product (for $x\neq 0$) is a superposition of two same-energy states $\ket{E}$ belonging to one of the combined subspaces $S=Ss+Sa$ or $A=As+As$,
\beq
\label{eq:superposition}
\ket{E,\pm x}_\gamma = \frac{1}{\sqrt{2}}(\ket{E}_{\gamma s}\pm \ket{E}_{\gamma a})~,\ \ \ \gamma=A,S
\eeq
The special case of $x=0$ consists of the states 
\beq
\ket{E,0}_S = \ket{E}_{Ss}~.
\eeq
The spectral properties of all four subspaces are similar, and hence the dynamics generated by them will also be similar. We consider only eigenstates belonging to the combined $A=As+As$ subspace, thus avoiding the complications induced by the $\ket{E,0}$ states. 

For $v=0$ both the direct products $\ket{E_0,\pm x_0}$ and their superpositions $\ket{E_0,x_0}\pm \ket{E_0,-x_0}$ are energy eigenstates. For a finite, but very small $v$ the spectrum consists of pairs of near-degenerate states that evolve from these two superpositions. In reverse, for a sufficiently small $v$ the states $\ket{E_0,\pm x_0}$ are mainly combinations of two near-degenerate eigenstates, and hence the flow of probability from $x_0$ to $-x_0$ is extremely slow. Even for very long simulation times we may observe the probability distribution spreading over a wide range of $x$ yet not crossing over to the other side of the $x$ axis. This creates two distinct time scales: one for spreading (by diffusion), and the other for saturation (by tunneling).

%%%%%%%%%%%%%%%%%%%%%%%%%%%%%%%%%%%%%%%%%%%%%%%%%%%%%%%%%%%%%%%%%%%%%%%%%%%%%%%%%%%%%%%%%%%%%%%%%%%%%%%%%%%%%%%%

%%%%%%%%%%%%%%%%%%%%%%%%%%%%%%%%%%%%%%%%%%%%%%%%%%%%%%%%%%%%%%%%%%%%
%%%%%%%%%%%%%%%%%%%%%%%%%%%%%%%%%%%%%%%%%%%%%%%%%%%%%%%%%%%%%%%%%%%%
\clearpage
\end{document}